\documentclass[aps,prd,onecolumn,preprintnumbers,superscriptaddress,tightenlines,amsmath,amssymb,showpacs,nofootinbib]{revtex4-2}
\usepackage{textcomp}
\usepackage[english]{babel}
\usepackage{amsmath,amssymb,amsbsy,booktabs}
\usepackage{bm}
\usepackage{xcolor}
\usepackage{array}
\usepackage{amstext}
\usepackage{graphicx}
\usepackage{amsfonts}
\usepackage{bm}
\usepackage{dcolumn}
\usepackage{rotating}
\usepackage{epstopdf}
\usepackage{esint}
\usepackage{url}
\usepackage{pifont}
\usepackage{colortbl}
\usepackage{longtable}
\usepackage[colorlinks=true,
  linkcolor=blue,
  filecolor=blue,
  anchorcolor=blue,
  urlcolor=blue,
  citecolor=blue
]{hyperref}

\usepackage[utf8]{inputenc}

\usepackage[T1]{fontenc} 

\usepackage{multirow}
\usepackage{makecell}

\usepackage{array}
\usepackage{booktabs}
\usepackage{times}
\usepackage{diagbox}
\usepackage{slashed}
\usepackage{simpler-wick}

\allowdisplaybreaks[0]

\begin{document}

\title {Probing strange-sector flavor-changing neutral currents at DUNE-like facilities}

\author{Xin-Shuai Yan}
\email{yanxinshuai@htu.edu.cn}
\affiliation{Institute of Particle and Nuclear Physics, Henan Normal University, Xinxiang, Henan 453007, China}

\author{Miao Liu}
\email{liumiao20239@stu.htu.edu.cn}
\affiliation{Institute of Particle and Nuclear Physics, Henan Normal University, Xinxiang, Henan 453007, China}

\author{Qin Chang}
\email{changqin@htu.edu.cn}
\affiliation{Institute of Particle and Nuclear Physics, Henan Normal University, Xinxiang, Henan 453007, China}

\author{Ya-Dong Yang}
\email{yangyd@mail.ccnu.edu.cn}
\affiliation{Institute of Particle and Nuclear Physics, Henan Normal University, Xinxiang, Henan 453007, China}
\affiliation{Institute of Particle Physics and Key Laboratory of Quark and Lepton Physics~(MOE),\\
Central China Normal University, Wuhan, Hubei 430079, China}

\begin{abstract}
  We investigate the sensitivity of DUNE-like facilities to flavor-changing
  neutral-current interactions between down and strange quarks through inclusive
  deep-inelastic neutrino scattering. In the framework of a general low-energy
  effective Lagrangian for Dirac and Majorana neutrinos, we evaluate projected
  sensitivities for CP- and $\tau$-optimized beam configurations. The strongest
  bounds on the dimensionless Wilson coefficients, defined relative to the Fermi
  constant $G_F$, reach $\mathcal{O}(10^{-3})\text{--}\mathcal{O}(10^{-2})$ at the
  near detector, whereas far-detector limits are roughly one order of magnitude
  weaker. When systematic uncertainties are included, the near-detector
  sensitivity is dominated by the background-rate uncertainty, while the
  far-detector sensitivity is limited primarily by the signal-rate uncertainty.
  At the far detector,
  the dependence of the projected bounds on the neutrino mass ordering and the Dirac
  CP-violating phase $\delta_{\mathrm{CP}}$ is largely confined to coefficients
  involving the electron flavor.
\end{abstract}

\maketitle

\section{Introduction}
\label{sec:intro}

Flavor-changing neutral-current (FCNC) processes are among the most sensitive probes of
physics beyond the Standard Model (SM).
They are forbidden at tree level in the SM and generated only through loop amplitudes, which are
further suppressed by the Glashow--Iliopoulos--Maiani (GIM) mechanism and the
hierarchical structure of the Cabibbo--Kobayashi--Maskawa matrix. Owing to this
suppression, FCNC processes can probe new-physics (NP) scales well
beyond the direct reach of collider experiments~\cite{Isidori:2010kg}.

FCNC transitions involving a neutrino pair, such as the rare $B$- and $K$-meson
decays, offer theoretically clean probes of NP~\cite{Isidori:2014rba,Buras:2013ooa}.
Such processes have recently drawn widespread interest
following the first evidence for $B^+\to K^+\nu\bar\nu$ reported by the Belle~II Collaboration,
with a branching fraction $2.7\sigma$ above the SM
prediction~\cite{Belle-II:2023esi}. Although the connection between
the bottom-sector $b\to s\nu\bar\nu$ transition and its strange-sector counterpart
$s\to d\nu\bar\nu$ is model dependent, NP models addressing
the Belle~II excess often predict correlated effects in $s\to d\nu\bar\nu$
transitions~\cite{Chen:2024cll,Marzocca:2024hua,Buras:2024ewl,Allwicher:2024ncl,Chen:2025npb,Liu:2025lbw,Hong:2026qoj,Abada:2026dlb}, strongly motivating dedicated studies of strange-sector dineutrino FCNC
processes.

The principal probes of $s\to d\nu\bar\nu$ transitions are the rare decays
$K^+\to\pi^+\nu\bar\nu$ and $K_L\to\pi^0\nu\bar\nu$. Recently, NA62 reported
$\mathcal{B}(K^+\to\pi^+\nu\bar\nu) = (9.6^{+1.9}_{-1.8})\times10^{-11}$,
consistent with the SM prediction~\cite{NA62:2026measurement}, while the neutral
mode remains unobserved, with the current KOTO upper limit
$\mathcal{B}(K_L\to\pi^0\nu\bar\nu)<2.2\times10^{-9}$ at 90\% confidence level
(C.L.)~\cite{KOTO:2024search}. The fully invisible two-body decays,
$K_{S,L}\to\text{invisible}$, provide complementary probes. BESIII recently
established the first direct upper limit,
$\mathcal{B}(K_S\to\text{invisible})<8.4\times 10^{-4}$ at 90\% C.L.~\cite{BESIII:2025invisible},
whereas no direct experimental measurement has yet been reported for $K_L\to\text{invisible}$.

In the absence of evidence for a specific ultraviolet completion, NP contributions to
$s\to d\nu\bar\nu$ can be described model independently within a low-energy effective field theory,
containing four-fermion operators with general Lorentz and chiral structures. Since the fundamental
nature of neutrinos remains unknown, both Dirac and Majorana hypotheses shall be
accommodated; while further possibilities such as mixed Dirac--Majorana schemes
also exist, the primary physical effects are well captured by these two benchmark
frameworks, which lead to distinct operator bases and a different number of independent
Wilson coefficients~\cite{Bischer:2019ttk,Gorbahn:2023juq}.
These rare $K$-meson decays constrain combinations of the Wilson coefficients but
cannot determine the complete operator basis. In particular, the differential distributions of
$K^+\to\pi^+\nu\bar\nu$ can distinguish certain Lorentz structures only under
restricted operator assumptions~\cite{Li:2019fhz, Gorbahn:2023juq}, while the invisible
two-body modes probe only a limited subset of interactions. Furthermore, these
decays cannot distinguish among neutrino flavors. Additional experimental
channels are therefore necessary.

By crossing symmetry, the same four-fermion operators mediate
flavor-changing neutrino--quark scattering. Since accelerator neutrino beams
provide an incident flux of known flavor composition that undergoes
oscillations along the baseline, inclusive neutral-current (NC) deep-inelastic
scattering (DIS) offers a direct probe of Wilson coefficients in the neutrino
flavor basis. At long-baseline facilities such as the Deep Underground
Neutrino Experiment (DUNE), high-intensity beams provide high-statistics
samples at the near detector (ND), while oscillations yield a different flavor
composition at the far detector (FD), enabling complementary sensitivity to
different neutrino-flavor combinations~\cite{DUNE:2020ypp,DUNE:2021cuw,DUNE:2024wvj}.
Previous work has demonstrated that this setup provides strong sensitivity to the
axial NC nonstandard neutrino interactions involving flavor-diagonal $u$, $d$, and $s$
quark currents~\cite{Abbaslu:2023vqk}.

In this work, we extend this approach to flavor-changing transitions between
down and strange quarks ($d\leftrightarrow s$), assessing the sensitivity of
inclusive NC DIS at DUNE-like facilities for both Dirac and Majorana
neutrinos. Neglecting the strongly GIM-suppressed SM contribution, we show
that the intense $\nu_\mu$ and $\bar{\nu}_\mu$ fluxes at the ND yield the
strongest constraints on Wilson coefficients with at least one muon-flavor
index.
Neutrino oscillations along the baseline broaden the flavor combinations
accessible at the FD, although the resulting limits are typically about an
order of magnitude weaker than those at the ND. While these scattering bounds
are numerically less stringent than indirect limits from
$K^+\to\pi^+\nu\bar\nu$~\cite{Gorbahn:2023juq}, they provide complementary
constraints that directly probe the neutrino flavor basis rather than the
mass basis. They also exhibit a sensitivity hierarchy opposite to that of the rare
kaon decay: tensor interactions are generally constrained most tightly,
whereas scalar and Majorana pseudoscalar interactions are constrained least
tightly. Furthermore, we examine the impact of
systematic uncertainties and oscillation parameters, demonstrating that the ND
sensitivity is particularly affected by background estimation uncertainty, whereas the FD sensitivity is limited
primarily by signal efficiency uncertainty and, for electron-flavor couplings, can depend
appreciably on the neutrino mass ordering and the Dirac CP-violating phase
$\delta_{\mathrm{CP}}$.

The remainder of this paper is organized as follows. In
Sec.~\ref{sec:framework}, we introduce the low-energy effective Lagrangians and
operator bases for $d\leftrightarrow s$ transitions involving Dirac and
Majorana neutrinos. In Sec.~\ref{sec:DIS}, we derive the inclusive NC DIS cross
sections and describe the event-rate calculation, beam configurations, and
neutrino flavor evolution. In Sec.~\ref{sec:sensitivity}, we present the
projected single-coefficient sensitivities, assess the effects of systematic
uncertainties and oscillation parameters, and examine selected two-coefficient
correlations. We summarize our conclusions in Sec.~\ref{sec:Summary}.

\section{Low-Energy Effective Lagrangian}
\label{sec:framework}

In the Dirac neutrino scenario, the general dimension-6 low-energy effective Lagrangian mediating the $s\to d \nu \bar{\nu}$ transition is given by
\begin{align}\label{eq:L6D}
  \mathcal{L}^{(6)}_D =  -\frac{G_F}{\sqrt{2}}\sum_{a,b}\sum_{X} C_{ab}^{X} O_{ab}^{X} + \text{h.c.} \,,
\end{align}
where $a, b \in \{e, \mu, \tau\}$ are flavor indices, $X \in \{V, S, T\}$, and the effective operators can be compactly written using chiral projectors ($\lambda, \kappa \in \{L, R\}$) as
\begin{align}
  O_{ab}^{V, \lambda\kappa}  & = (\bar{\nu}_{a}\gamma_\mu P_\lambda \nu_{b})(\bar{d} \gamma^\mu P_\kappa s)\,, \nonumber   \\[0.2cm]
  O_{ab}^{S, \lambda\kappa}  & = (\bar{\nu}_{a} P_\lambda \nu_{b})(\bar{d} P_\kappa s)\,, \nonumber                        \\[0.2cm]
  O_{ab}^{T, \lambda\lambda} & = (\bar{\nu}_{a} \sigma_{\mu\nu} P_\lambda \nu_{b})(\bar{d} \sigma^{\mu\nu} P_\lambda s)\,.
\end{align}
Note that tensor operators with mixed chiralities ($\lambda \neq \kappa$) vanish identically, and lepton number is
conserved in this case. The Wilson coefficients $C_{ab}^{X}$ are normalized relative to the Fermi constant $G_F$ to render them dimensionless and to facilitate a direct comparison with the SM weak interaction strength. Accounting for all possible flavor combinations, this basis yields 90 independent operators~\cite{Bischer:2019ttk, Gorbahn:2023juq}.

In the Majorana neutrino scenario, to distinguish from the Dirac case, we denote the effective operators by $\mathcal{O}$. The corresponding dimension-6 effective Lagrangian is given by
\begin{align}\label{eq:M6D}
  \mathcal{L}^{(6)}_M =  -\frac{G_F}{\sqrt{2}}\sum_{a\geq b}\sum_{X, \kappa} \tilde{C}_{ab}^{X,\kappa} \mathcal{O}_{ab}^{X,\kappa} + \text{h.c.} \,,
\end{align}
where $X \in \{V, A, S, P, T\}$ and $\kappa \in \{L, R\}$. The specific operators are defined as
\begin{align}
  \mathcal{O}_{ab}^{V, \kappa} & = \tfrac{1}{2}(\bar{\nu}_{a} \gamma_\mu \nu_{b})(\bar{d} \gamma^\mu P_\kappa s)\,,                    &
  \mathcal{O}_{ab}^{A, \kappa} & = \tfrac{1}{2}(\bar{\nu}_{a} \gamma_\mu \gamma_5 \nu_{b})(\bar{d} \gamma^\mu P_\kappa s)\,, \nonumber   \\[0.2cm]
  \mathcal{O}_{ab}^{S, \kappa} & = \tfrac{1}{2}(\bar{\nu}_{a} \nu_{b})(\bar{d} P_\kappa s)\,,                                          &
  \mathcal{O}_{ab}^{P, \kappa} & = \tfrac{1}{2}(\bar{\nu}_{a} i\gamma_5 \nu_{b})(\bar{d} P_\kappa s)\,, \nonumber                        \\[0.2cm]
  \mathcal{O}_{ab}^{T, \kappa} & = \tfrac{1}{2}(\bar{\nu}_{a} \sigma_{\mu\nu} \nu_{b})(\bar{d} \sigma^{\mu\nu} P_\kappa s)\,.
\end{align}
By construction, the Wilson coefficients in this scenario exhibit the symmetry property $\tilde{C}_{ab}^{X,\kappa} = \eta_X \tilde{C}_{ba}^{X,\kappa}$, where
\begin{equation}
  \eta_X =
  \begin{cases}
    +1 & \text{for } X \in \{A, S, P\} \\
    -1 & \text{for } X \in \{V, T\}
  \end{cases}\,. \label{eq:symmetry}
\end{equation}
Consequently, the flavor-diagonal vector and tensor coefficients vanish
identically ($\tilde{C}_{aa}^{V,\kappa} = \tilde{C}_{aa}^{T,\kappa} = 0$),
thereby reducing the basis to 48 independent operators~\cite{Gorbahn:2023juq}.

While further possibilities exist—such as mixed Dirac--Majorana states or Dirac
neutrinos with lepton-number-violating dimension-6 interactions—the essential
physical effects can be fully illustrated within the two benchmark cases
described above.
\section{Neutral-Current DIS Framework}
\label{sec:DIS}

The corresponding neutrino--nucleon DIS observables arise from
flavor-changing partonic processes initiated by $d$, $s$, $\bar d$, or
$\bar s$ constituents of the nucleon. The Wilson coefficients therefore enter
the inclusive NC cross sections through
convolutions with the relevant parton distribution functions (PDFs). In this section,
we first set up the DIS kinematics and then derive the Dirac and Majorana NP
contributions to the cross sections used in the phenomenological
analysis.

\subsection{Kinematics}

We consider the following inclusive NC DIS processes for neutrinos and
antineutrinos on nucleon targets:
\begin{align}
  \nu_a(p_1)+N(p_2)       & \to \nu_b(p_3)+X,          \\[0.2cm]
  \bar{\nu}_a(p_1)+N(p_2) & \to \bar{\nu}_b(p_3)+X, 
\end{align}
where $N\in\{p,n\}$ is the target nucleon and $X$ denotes the inclusive
hadronic final state.

In the laboratory frame, where the target nucleon of mass $M$ is at rest, the
relevant four-momenta are chosen as
\begin{equation}
  p_1^\mu=(E_\nu,\vec{p}_1), \qquad
  p_2^\mu=(M,\vec{0}), \qquad
  p_3^\mu=(E'_\nu,\vec{p}_3)\,.
\end{equation}
Since the momentum transfers relevant for DIS are much larger than the neutrino
masses, neutrinos are treated as massless, so that
$|\vec{p}_1|=E_\nu$ and $|\vec{p}_3|=E'_\nu$. With $q=p_1-p_3$, we use the
standard DIS variables: the spacelike momentum transfer $Q^2\equiv -q^2>0$,
the Bjorken scaling variable $x$, and the inelasticity $y$, defined by
\begin{align}
  x & = \frac{Q^2}{2p_2\cdot q}
  = \frac{Q^2}{2M(E_\nu-E'_\nu)}\,, \\[0.2cm]
  y & = \frac{p_2\cdot q}{p_2\cdot p_1}
  = 1-\frac{E'_\nu}{E_\nu}\,.
\end{align}
The physical region is then $0\leq x\leq 1$ and
$0\leq y\leq y_{\mathrm{max}}$, with
\begin{equation}
  y_{\mathrm{max}}=\left(1+\frac{Mx}{2E_\nu}\right)^{-1}\,.
\end{equation}

In the collinear parton model, the double-differential cross section for a
partonic transition $\nu_a q_i\to \nu_b q_j$ on a target nucleon is
obtained by convolving the partonic squared matrix element with the
corresponding target PDF. For massless partons, it is given by~\cite{Liu:2015rqa}
\begin{equation}
  \frac{d^2\sigma_N}{dx\,dy}
  =\frac{1}{32\pi M E_\nu}\int_0^1\frac{d\xi}{\xi}\,
  f_N^{q_i}(\xi,Q^2)\,
  \overline{|\mathcal{M}(\xi)|^2}_{q_i\to q_j}\,\delta(\xi-x)\,.
\end{equation}
Here $f_N^{q_i}(\xi,Q^2)$ is the PDF of the initial-state quark or antiquark of
flavor $q_i$ in the target nucleon $N$, evaluated at the factorization scale
$Q^2$, and $\overline{|\mathcal{M}|^2}$ denotes the spin-averaged squared matrix
element for the underlying partonic process. The delta function fixes the
parton momentum fraction to the Bjorken variable, $\xi=x$.

For proton targets, we use the proton PDFs directly. The neutron PDFs are
obtained by isospin symmetry: the up-quark distribution in the neutron is
identified with the down-quark distribution in the proton, and vice versa,
with the same interchange applied to the corresponding antiquark
distributions. The strange and antistrange distributions are unchanged.
For numerical analysis, we employ the CT18 next-to-next-to-leading order
(NNLO) PDF sets~\cite{Hou:2019efy, Carrazza:2014gfa, Bertone:2013vaa}. Since
the momentum transfers relevant for DUNE-like accelerator beams are concentrated
at relatively low scales, $Q^2\lesssim 10~\mathrm{GeV}^2$, and the scaling
violations over this range are mild, the PDFs are evaluated at the fixed
reference scale $Q=2~\mathrm{GeV}$, as in Ref.~\cite{Abbaslu:2023vqk}. At this
scale, the PDF moments entering the cross sections are listed in
Table~\ref{tab:quark_moments}; in the operator-specific expressions below, the
target label is omitted where unambiguous.

\begin{table*}[t]
  \centering
  \renewcommand{\arraystretch}{1.5}
  \setlength{\tabcolsep}{35pt}
  \caption{Moments of the quark and antiquark distribution functions, $\int_0^1 dx\, x^n f^{q}_p(x)$ and $\int_0^1 dx\, x^n f^{\bar{q}}_p(x)$ for $n=1, 2, 3$. The integrals are evaluated for the $u$, $d$, and $s$ quarks at a fixed reference scale of $Q = 2~\mathrm{GeV}$ using the CT18 NNLO PDF sets~\cite{Hou:2019efy, Carrazza:2014gfa, Bertone:2013vaa}.}
  \begin{tabular}{lccc}
    \toprule
    \toprule
    Integral & $u$                & $d$                & $s$                \\
    \midrule
    $\int_0^1 dx\, x\, f^{q}_p(x)$
    & $0.320 \pm 0.0043$ & $0.157 \pm 0.0040$ & $0.016 \pm 0.0046$ \\
    $\int_0^1 dx\, x^2\, f^{q}_p(x)$
    & $0.087 \pm 0.0013$ & $0.034 \pm 0.0014$ & $0.001 \pm 0.0005$ \\
    $\int_0^1 dx\, x^3\, f^{q}_p(x)$
    & $0.033 \pm 0.0006$ & $0.011 \pm 0.0008$ & $0.000 \pm 0.0001$ \\
    \midrule
    $\int_0^1 dx\, x\, f^{\bar{q}}_p(x)$
    & $0.030 \pm 0.0020$ & $0.037 \pm 0.0025$ & $0.016 \pm 0.0046$ \\
    $\int_0^1 dx\, x^2\, f^{\bar{q}}_p(x)$
    & $0.003 \pm 0.0003$ & $0.004 \pm 0.0004$ & $0.001 \pm 0.0005$ \\
    $\int_0^1 dx\, x^3\, f^{\bar q}_p(x)$
    & $0.001 \pm 0.0001$ & $0.001 \pm 0.0001$ & $0.000 \pm 0.0001$ \\
    \bottomrule
    \bottomrule
  \end{tabular}
  \label{tab:quark_moments}
\end{table*}

The total cross section for each target nucleon, $\sigma_N$ with $N=p,n$, is
obtained by integrating the differential distribution over the allowed phase
space. Expanding the upper limit on the inelasticity in powers of
$\alpha\equiv M/(2E_\nu)$ gives
\begin{equation}
  y_{\mathrm{max}}=1-\alpha x+\alpha^2x^2+\mathcal{O}(\alpha^3)\,.
\end{equation}
For the neutrino energies relevant to DUNE-like facilities, truncating this
series at second order induces a fractional error of order $10^{-4}$, which is
negligible for the present analysis~\cite{Abbaslu:2023vqk}.

\vspace{-1.5em}

\subsection{Cross Sections for Dirac and Majorana Neutrinos}

Building on the general DIS cross-section formulation presented above, we now
detail the contributions from the different effective operators introduced in Sec.~\ref{sec:framework}.
For both the Dirac and Majorana scenarios, we consider neutrino and antineutrino initial
states separately.

\vspace{-1.5em}
\subsubsection{Dirac-Neutrino Cross Sections}
\label{sec:NP_Dirac}

We first consider the DIS cross sections generated by the Dirac-neutrino
operators. Although chirality and helicity are distinct for massive fermions,
they can be identified for the effectively massless neutrinos considered in
the scattering calculation. We therefore label the external states by
chirality. The incident beam is assumed to be produced through SM
charged-current interactions and thus consists of left-handed neutrinos or
right-handed antineutrinos. For the flavor-changing transition in
the strange--down sector, the relevant initial partons are
$q=s,d,\bar{s},\bar{d}$.

The partonic processes mediated by each operator are summarized in Table~\ref{tab:DIS_processes}.
For flavor-off-diagonal transitions ($a\neq b$), each operator generates two
partonic channels for each allowed incoming neutrino flavor. The vector
operators $O_{ab}^{V,L\kappa}$ admit  both incoming flavors: $\nu_b$ scatters
from $s$ and $\bar d$ partons, whereas $\nu_a$ scatters from $d$ and $\bar s$
partons. By contrast, scalar and tensor operators with fixed neutrino-current chirality
select only one incoming flavor. Specifically, $O_{ab}^{S,L\kappa}$ and
$O_{ab}^{T,LL}$ mediate scattering of an incoming $\nu_b$, whereas
$O_{ab}^{S,R\kappa}$ and $O_{ab}^{T,RR}$ mediate scattering of an incoming
$\nu_a$. In the
flavor-diagonal case ($a=b$), the two neutrino flavors coincide. Consequently,
all four partonic channels contribute to vector operators, whereas only two
contribute to scalar and tensor operators. The same pattern applies to the
corresponding right-handed antineutrino states.

\begin{table*}[t]
  \centering
  \renewcommand{\arraystretch}{1.2}
  \setlength{\tabcolsep}{50pt}
  \caption{Partonic DIS channels induced by the Dirac dimension-6 operators. The symbol $\kappa$ denotes quark chirality, $\kappa\in\{L,R\}$.}
  \begin{tabular}{l c c}
    \toprule
    \toprule
    \textbf{Operator} & \textbf{Initial State} & \textbf{Physical Process} \\
    \midrule

    \multirow{4}{*}{$O^{V,L\kappa}_{ab}$}
    & \multirow{2}{*}{$|\nu_L\rangle$}
    & $\nu_b s \to \nu_a d$, \quad $\nu_a d \to \nu_b s$ \\

    &
    & $\nu_b \bar{d} \to \nu_a \bar{s}$, \quad $\nu_a \bar{s} \to \nu_b \bar{d}$ \\

    \cmidrule{2-3}

    & \multirow{2}{*}{$|\bar{\nu}_L\rangle$}
    & $\bar{\nu}_a s \to \bar{\nu}_b d$, \quad $\bar{\nu}_b d \to \bar{\nu}_a s$ \\

    &
    & $\bar{\nu}_a \bar{d} \to \bar{\nu}_b \bar{s}$, \quad $\bar{\nu}_b \bar{s} \to \bar{\nu}_a \bar{d}$ \\

    \midrule

    $O^{S,L\kappa}_{ab}$, $O^{T,LL}_{ab}$
    & $|\nu_L\rangle$
    & $\nu_b s \to \nu_a d$, \quad $\nu_b \bar{d} \to \nu_a \bar{s}$ \\

    $O^{S,R\kappa}_{ab}$, $O^{T,RR}_{ab}$
    & $|\nu_L\rangle$
    & $\nu_a d \to \nu_b s$, \quad $\nu_a \bar{s} \to \nu_b \bar{d}$ \\

    \midrule

    $O^{S,L\kappa}_{ab}$, $O^{T,LL}_{ab}$
    & $|\bar{\nu}_L\rangle$
    & $\bar{\nu}_b d \to \bar{\nu}_a s$, \quad $\bar{\nu}_b \bar{s} \to \bar{\nu}_a \bar{d}$ \\

    $O^{S,R\kappa}_{ab}$, $O^{T,RR}_{ab}$
    & $|\bar{\nu}_L\rangle$
    & $\bar{\nu}_a s \to \bar{\nu}_b d$, \quad $\bar{\nu}_a \bar{d} \to \bar{\nu}_b \bar{s}$ \\

    \bottomrule
  \end{tabular}
  \label{tab:DIS_processes}
\end{table*}

The Dirac contributions can then be written compactly for the four possible
incoming states, $i\in\{\nu_b,\nu_a,\bar\nu_a,\bar\nu_b\}$.
At leading order in the parton model, the corresponding double-differential
cross section reads
\begin{equation}
  \label{eq:diff_cross_section}
  \begin{aligned}
    \frac{d^2\sigma_{i}}{dx dy} = \frac{G^2_F E_{\nu} M x}{32\pi} \bigg\{
      &4 \left[ f_1^{i}(x, Q^2) + (1-y)^2 f_2^{i}(x, Q^2) \right] |C_{ab}^{V, LL}|^{2} \\[0.2cm]
      +\, &4 \left[ (1-y)^2 f_1^{i}(x, Q^2) + f_2^{i}(x, Q^2) \right] |C_{ab}^{V, LR}|^{2} \\[0.2cm]
      +\, &y^2 \left[ f_1^{i}(x, Q^2) + f_2^{i}(x, Q^2) \right] \left( |C_{ab}^{S, \chi_i L}|^{2} + |C_{ab}^{S, \chi_i R}|^{2} \right) \\[0.2cm]
      -\, &8 y(2-y) \left[ f_1^{i}(x, Q^2) - f_2^{i}(x, Q^2) \right] \Re\left[C_{ab}^{S, \chi_i \chi_i} C_{ab}^{T, \chi_i \chi_i*}\right] \\[0.2cm]
    +\, &16 (2-y)^2 \left[ f_1^{i}(x, Q^2) + f_2^{i}(x, Q^2) \right] |C_{ab}^{T, \chi_i \chi_i}|^{2} \bigg\}\,.
  \end{aligned}
\end{equation}
Here $\chi_i$ denotes the neutrino-current chirality selected by the scalar and
tensor coefficients: $\chi_i=L$ for $i\in\{\nu_b,\bar\nu_b\}$ and $\chi_i=R$
for $i\in\{\nu_a,\bar\nu_a\}$. The functions $f_1^i$ and $f_2^i$ denote the
relevant quark and antiquark PDFs for each incoming state and are defined by
\begin{equation}
  \begin{array}{c|cccc}
    i & \nu_b & \nu_a & \bar\nu_a & \bar\nu_b \\
    \hline  \\[-1em]
    (f_1^i,f_2^i) & (f^s,f^{\bar d}) & (f^d,f^{\bar s}) & (f^{\bar d},f^s) & (f^{\bar s},f^d)
  \end{array} \nonumber
\end{equation}
where each PDF in the table is understood to be evaluated at $(x,Q^2)$.

The inclusive cross section follows by integrating Eq.~\eqref{eq:diff_cross_section} over $0\leq x\leq1$ and $0\leq y\leq y_{\rm max}$. As described above, we expand $y_{\rm max}$ in powers of $\alpha=M/(2E_\nu)$ and retain terms through $\mathcal{O}(\alpha^2)$. Since the PDFs are evaluated at the fixed reference scale $Q=2~\mathrm{GeV}$, their scale dependence is suppressed and only the remaining $x$ integration is kept explicit. This leads to
\begin{equation}
  \label{eq:total_cross_section}
  \begin{aligned}
    \sigma_{i} = \frac{G^2_F E_{\nu} M }{32\pi} \int_{0}^{1} dx \, x \bigg\{
      &4 \left[ (1 - x \alpha + x^2 \alpha^2) f_1^{i}(x)  + \frac{1}{3} f_2^{i}(x) \right] |C_{ab}^{V, LL}|^{2} \\[0.2cm]
      +\, &4 \left[ \frac{1}{3} f_1^{i}(x) + (1 - x \alpha + x^2 \alpha^2) f_2^{i}(x) \right] |C_{ab}^{V, LR}|^{2} \\[0.2cm]
      +\, &\left(\frac{1}{3} - x \alpha + 2 x^2 \alpha^2\right) \left[ f_1^{i}(x) + f_2^{i}(x) \right] \left( |C_{ab}^{S, \chi_i L}|^{2} + |C_{ab}^{S, \chi_i R}|^{2} \right) \\[0.2cm]
      -\, &8 \left(\frac{2}{3} - x \alpha + x^2 \alpha^2\right) \left[ f_1^{i}(x) - f_2^{i}(x) \right] \Re\left[C_{ab}^{S, \chi_i \chi_i}C_{ab}^{T, \chi_i \chi_i*}\right] \\[0.2cm]
    +\, &16 \left(\frac{7}{3} - x \alpha\right) \left[ f_1^{i}(x) + f_2^{i}(x) \right] |C_{ab}^{T, \chi_i \chi_i}|^{2} \bigg\} \,.
  \end{aligned}
\end{equation}

\subsubsection{Majorana-Neutrino Cross Sections}
\label{Sec:Majorana_NP}

We next consider the DIS cross sections induced by the Majorana-neutrino
operators. Since a Majorana neutrino is
identical to its charge conjugate, the states identified as
antineutrinos in SM weak decays may equivalently be described as right-handed
states, $|\nu_R\rangle$. It is therefore convenient to label the
incoming states by flavor and chirality rather than by neutrino versus
antineutrino beams. We consider the four possible incoming configurations
$\nu_{La}$, $\nu_{Lb}$, $\nu_{Ra}$, and $\nu_{Rb}$.

To write the results compactly, we introduce a generic incoming state
$\nu_{h\alpha}$, where $h\in\{L,R\}$ denotes the initial chirality and
$\alpha\in\{a,b\}$ denotes the initial flavor. We also define the sign factors
$s_h=+1(-1)$ for $h=R(L)$ and $s_\alpha=+1(-1)$ for $\alpha=a(b)$.
The opposite chirality is denoted by $\bar h$, with $\bar L=R$ and $\bar R=L$.

In the flavor-off-diagonal case ($a \neq b$), the double-differential cross
section for an incoming $\nu_{h\alpha}$ is given by
\begin{align}
  \label{eq:M_cro_sec1}
  \frac{d^2\sigma_{\nu_{h\alpha}}}{dx dy} = \frac{G^2_F E_{\nu} M x}{128\pi} \bigg\{
    &4 \left[ f^{s}(x, Q^2) + (1-y)^2 f^{\bar{d}}(x, Q^2) \right] |\tilde{C}_{ab}^{V, h} - s_h s_\alpha \tilde{C}_{ab}^{A, h}|^{2} \nonumber \\[0.1cm]
    +\, &4 \left[ f^{d}(x, Q^2) + (1-y)^2 f^{\bar{s}}(x, Q^2) \right] |\tilde{C}_{ab}^{V, h} + s_h s_\alpha \tilde{C}_{ab}^{A, h}|^{2} \nonumber \\[0.1cm]
    +\, &4 \left[ (1-y)^2f^{s}(x, Q^2) +  f^{\bar{d}}(x, Q^2) \right] |\tilde{C}_{ab}^{V, \bar{h}} - s_h s_\alpha \tilde{C}_{ab}^{A, \bar{h}}|^{2} \nonumber \\[0.1cm]
    +\, &4 \left[ (1-y)^2f^{d}(x, Q^2) +  f^{\bar{s}}(x, Q^2) \right] |\tilde{C}_{ab}^{V, \bar{h}} + s_h s_\alpha \tilde{C}_{ab}^{A, \bar{h}}|^{2} \nonumber \\[0.1cm]
    +\, &y^2 \sum_{\kappa \in \{L,R\}} \left[ f^{s}(x, Q^2) + f^{\bar{d}}(x, Q^2)+ f^{d}(x, Q^2) + f^{\bar{s}}(x, Q^2) \right] \left| \tilde{C}_{ab}^{S, \kappa} + i s_h \tilde{C}_{ab}^{P, \kappa} \right|^{2} \nonumber \\[0.1cm]
    +\, &8 s_\alpha y(2-y) \left[ f^{s}(x, Q^2) - f^{\bar{d}}(x, Q^2) \right] \Re\left[(\tilde{C}_{ab}^{S, h} + i s_h \tilde{C}_{ab}^{P, h})\tilde{C}_{ab}^{T, h*}\right] \nonumber \\[0.1cm]
    -\, &8 s_\alpha y(2-y) \left[ f^{d}(x, Q^2) - f^{\bar{s}}(x, Q^2) \right] \Re\left[(\tilde{C}_{ab}^{S, \bar{h}} + i s_h \tilde{C}_{ab}^{P, \bar{h}})\tilde{C}_{ab}^{T, \bar{h}*}\right] \nonumber \\[0.1cm]
    +\, &16 (2-y)^2 \left[ f^{s}(x, Q^2) + f^{\bar{d}}(x, Q^2) \right] |\tilde{C}_{ab}^{T, h}|^{2}\nonumber \\[0.1cm]
  +\, &16 (2-y)^2 \left[ f^{d}(x, Q^2) + f^{\bar{s}}(x, Q^2) \right] |\tilde{C}_{ab}^{T, \bar{h}}|^{2} \bigg\}.
\end{align}
Integrating over $y$ and retaining terms through $\mathcal{O}(\alpha^2)$ yields
the corresponding total cross section,
\begin{align}
  \label{eq:M_cro_sec2}
  \sigma_{\nu_{h\alpha}} = \frac{G^2_F E_{\nu} M }{128\pi} \int_{0}^{1} dx \, x  \bigg\{
    &4 \left[ (1 - x \alpha + x^2 \alpha^2)  f^{s}(x) + \frac{1}{3} f^{\bar{d}}(x) \right] |\tilde{C}_{ab}^{V, h} - s_h s_\alpha \tilde{C}_{ab}^{A, h}|^{2} \nonumber \\
    +\, &4 \left[(1 - x \alpha + x^2 \alpha^2)  f^{d}(x) +  \frac{1}{3} f^{\bar{s}}(x) \right] |\tilde{C}_{ab}^{V, h} + s_h s_\alpha \tilde{C}_{ab}^{A, h}|^{2} \nonumber \\
    +\, &4 \left[  \frac{1}{3}f^{s}(x) + (1 - x \alpha + x^2 \alpha^2) f^{\bar{d}}(x) \right] |\tilde{C}_{ab}^{V, \bar{h}} - s_h s_\alpha \tilde{C}_{ab}^{A, \bar{h}}|^{2} \nonumber \\
    +\, &4 \left[  \frac{1}{3} f^{d}(x) +(1 - x \alpha + x^2 \alpha^2)  f^{\bar{s}}(x) \right] |\tilde{C}_{ab}^{V, \bar{h}} + s_h s_\alpha \tilde{C}_{ab}^{A, \bar{h}}|^{2} \nonumber \\
    +\, &\left(\frac{1}{3} - x \alpha + 2 x^2 \alpha^2\right) \sum_{\kappa \in \{L,R\}} \left[ f^{s}(x) + f^{\bar{d}}(x)+ f^{d}(x) + f^{\bar{s}}(x) \right] \left| \tilde{C}_{ab}^{S, \kappa} + i s_h \tilde{C}_{ab}^{P, \kappa} \right|^{2} \nonumber \\
    +\, &8 s_\alpha  \left(\frac{2}{3} - x \alpha + x^2 \alpha^2\right) \left[ f^{s}(x) - f^{\bar{d}}(x) \right] \Re\left[(\tilde{C}_{ab}^{S, h} + i s_h \tilde{C}_{ab}^{P, h})\tilde{C}_{ab}^{T, h*}\right] \nonumber \\
    -\, &8 s_\alpha  \left(\frac{2}{3} - x \alpha + x^2 \alpha^2\right) \left[ f^{d}(x) - f^{\bar{s}}(x) \right] \Re\left[(\tilde{C}_{ab}^{S, \bar{h}} + i s_h \tilde{C}_{ab}^{P, \bar{h}})\tilde{C}_{ab}^{T, \bar{h}*}\right] \nonumber \\
    +\, &16 \left(\frac{7}{3} - x \alpha\right)  \left[ f^{s}(x) + f^{\bar{d}}(x) \right] |\tilde{C}_{ab}^{T, h}|^{2}\nonumber \\
  +\, &16  \left(\frac{7}{3} - x \alpha\right)  \left[ f^{d}(x) + f^{\bar{s}}(x) \right] |\tilde{C}_{ab}^{T, \bar{h}}|^{2} \bigg\}.
\end{align}

Eqs.~\eqref{eq:M_cro_sec1} and \eqref{eq:M_cro_sec2} show that, for a fixed
incoming Majorana state, the vector, axial-vector, scalar, and pseudoscalar
operators can mediate all four partonic processes. For instance,
an incoming $\nu_b$ of fixed chirality can participate in $\nu_b s \to \nu_a d$, $\nu_b d \to \nu_a s$,
$\nu_b \bar d \to \nu_a \bar s$, and $\nu_b \bar s \to \nu_a \bar d$.
This differs from the Dirac case, where the effective operators mediate only
two of these channels for a state of fixed chirality. This enlarged set of processes follows from the Majorana
condition, $\psi=\psi^c$. Tensor operators, however, are an exception. For a fixed incoming chirality,
only two of the four channels contribute, because tensor operators with mixed
chiralities vanish. Given that the possible Majorana partonic processes have been fully
specified above, we do not provide a separate table analogous to
Table~\ref{tab:DIS_processes}.

For flavor-diagonal transitions ($a=b$), the vector and tensor coefficients
vanish identically, $\tilde{C}_{aa}^{V,\kappa}=\tilde{C}_{aa}^{T,\kappa}=0$.
The double-differential cross section for an incoming $\nu_{ha}$ then reduces to
\begin{align}
  \frac{d^2\sigma_{\nu_{h a}}}{dx dy} = \frac{G^2_F E_{\nu} M x}{32\pi}\bigg\{
    &4 \left[ \left(f^{s}(x, Q^2)+f^{d}(x, Q^2)\right) + (1-y)^2 \left(f^{\bar{d}}(x, Q^2)+f^{\bar{s}}(x, Q^2)\right)  \right] |\tilde{C}_{aa}^{A, h}|^{2} \nonumber \\[0.2cm]
    +\, &4 \left[ \left(f^{\bar{d}}(x, Q^2)+f^{\bar{s}}(x, Q^2)\right) + (1-y)^2\left(f^{s}(x, Q^2)+f^{d}(x, Q^2)\right) \right] |\tilde{C}_{aa}^{A, \bar{h}}|^{2} \nonumber \\[0.2cm]
  +\, &y^2 \sum_{\kappa \in \{L,R\}} \left[ f^{s}(x, Q^2) + f^{\bar{d}}(x, Q^2)+ f^{d}(x, Q^2) + f^{\bar{s}}(x, Q^2) \right] \left| \tilde{C}_{aa}^{S, \kappa} + i s_h \tilde{C}_{aa}^{P, \kappa} \right|^{2}  \bigg\}.
\end{align}
The overall factor of four relative to the flavor-off-diagonal case stems from the factor-of-two enhancement of the flavor-diagonal Majorana amplitude. The corresponding total cross section reads
\begin{align}
  \sigma_{\nu_{h a}} = \frac{G^2_F E_{\nu} M }{32\pi} \int_{0}^{1} dx \, x  \bigg\{
    &4 \left[ (1 - x \alpha + x^2 \alpha^2)  \left( f^{s}(x)+f^{d}(x)\right)  + \frac{1}{3} \left(f^{\bar{d}}(x)+f^{\bar{s}}(x)\right) \right] |\tilde{C}_{aa}^{A, h}|^{2} \nonumber \\[0.2cm]
    +\, &4\left[ (1 - x \alpha + x^2 \alpha^2) \left(f^{\bar{d}}(x)+f^{\bar{s}}(x)\right)  + \frac{1}{3}  \left( f^{s}(x)+f^{d}(x)\right)  \right] |\tilde{C}_{aa}^{A, \bar{h}}|^{2} \nonumber \\[0.2cm]
  +\, &\left(\frac{1}{3} - x \alpha + 2 x^2 \alpha^2\right) \sum_{\kappa \in \{L,R\}} \left[ f^{s}(x) + f^{\bar{d}}(x)+ f^{d}(x) + f^{\bar{s}}(x) \right] \left| \tilde{C}_{aa}^{S, \kappa} + i s_h \tilde{C}_{aa}^{P, \kappa} \right|^{2} \bigg\}.
\end{align}

\subsection{Standard Model Neutral-Current DIS}
\label{app:SM_cross_sections}

In the SM, neutrino NC scattering is mediated at tree level by
$Z$-boson exchange. At momentum transfers $Q^2 \ll M_Z^2$, the interaction is
described by the flavor-diagonal effective Lagrangian
\begin{align}
  \mathcal{L}_{\text{SM}}^{\text{NC}} = -\frac{4 G_F}{\sqrt{2}} \sum_{a} \left( \bar{\nu}_a \gamma^\mu P_L \nu_a \right) \sum_{q} \bar{q} \gamma_\mu \left( g_L^q P_L + g_R^q P_R \right) q \,.
\end{align}
Here $a \in \{e, \mu, \tau\}$, $q$ denotes an active quark flavor, and the chiral
couplings are given by
\begin{equation}
  g_L^q = T_3^q - Q_q \sin^2\theta_W\,, \qquad
  g_R^q = - Q_q \sin^2\theta_W\,.
\end{equation}
For the light quarks relevant to the present analysis,
\begin{align}
  g_L^u     & = \frac{1}{2} - \frac{2}{3}\sin^2\theta_W\,,  & g_R^u     & = -\frac{2}{3}\sin^2\theta_W\,, \nonumber \\[0.2cm]
  g_L^{d,s} & = -\frac{1}{2} + \frac{1}{3}\sin^2\theta_W\,, & g_R^{d,s} & = \frac{1}{3}\sin^2\theta_W\,.
\end{align}

The corresponding inclusive neutrino--nucleon DIS cross section takes the form
\begin{align}
  \frac{d^2\sigma_{\text{SM}}}{dx dy}(\nu_a N\to \nu_a X) = \frac{2G_F^2 M E_\nu}{\pi} x \sum_{q} \bigg\{
    &\Big[ (g_L^q)^2 + (g_R^q)^2 (1-y)^2 \Big] f^q(x,Q^2) \nonumber \\[0.2cm]
  +\,&\Big[ (g_R^q)^2 + (g_L^q)^2 (1-y)^2 \Big] f^{\bar{q}}(x,Q^2) \bigg\}\,.
\end{align}
For antineutrino scattering, the chirality-dependent kinematic factors are
interchanged
\begin{align}
  \frac{d^2\sigma_{\text{SM}}}{dx dy}(\bar{\nu}_{a} N\to \bar{\nu}_{a} X) = \frac{2G_F^2 M E_\nu}{\pi} x \sum_{q} \bigg\{
    &\Big[ (g_R^q)^2 + (g_L^q)^2 (1-y)^2 \Big] f^q(x,Q^2) \nonumber \\[0.2cm]
  +\,&\Big[ (g_L^q)^2 + (g_R^q)^2 (1-y)^2 \Big] f^{\bar{q}}(x,Q^2) \bigg\}\,.
\end{align}
These expressions are independent of the neutrino flavor $a$, reflecting the
flavor universality of the SM NC interaction. In the limit
$m_\nu/E_\nu \to 0$, they apply equally to Dirac and Majorana neutrinos.

Since the NP operators considered here induce $s \leftrightarrow d$
transitions, interference with the SM background would require corresponding
flavor-changing SM amplitudes. Such transitions are absent in the SM at tree
level and are strongly GIM-suppressed at loop level. Neglecting these suppressed
contributions, the SM background retained above is flavor-conserving,
precluding any interference with the flavor-changing NP signal. The total cross
section is therefore obtained by adding the NP signal incoherently to the SM
rate.

\subsection{Event Rate Formulation}
\label{sec:event_rate_formulation}

To convert the nucleon cross sections derived above into event
rates, we apply them to the $^{40}\mathrm{Ar}$ nuclear targets used in DUNE-like detectors.
We employ the free-nucleon approximation, treating the inclusive nuclear cross
section as an incoherent sum over the constituent protons and neutrons. Nuclear
medium effects---such as shadowing and the EMC effect---can induce corrections
at the $\mathcal{O} (10\%\text{--}20\%)$ level.
Nevertheless, the free-nucleon approximation is sufficient to estimate the sensitivity to
neutrino nonstandard interactions~\cite{Altmannshofer:2018xyo} and, analogously,
the FCNC processes of interest here.

For $^{40}\mathrm{Ar}$, neglecting binding energies and the proton--neutron mass
difference, the number of target protons and neutrons in a given detector is
\begin{equation}
  N_p^{\mathrm{ND/FD}} \simeq
  \frac{18}{40}\frac{M_{\mathrm{fid}}^{\mathrm{ND/FD}}}{M_p}
  \,,\qquad
  N_n^{\mathrm{ND/FD}} \simeq
  \frac{22}{40}\frac{M_{\mathrm{fid}}^{\mathrm{ND/FD}}}{M_p}\,,
\end{equation}
where $M_p$ denotes the proton mass. We use fiducial masses of $67.2~\text{t}$
for the ND and $40~\text{kt}$ for the FD~\cite{DUNE:2021cuw,DUNE:2020ypp}.

The predicted event yields further depend on the flavor composition of the
neutrino and antineutrino beams at each detector. Since DUNE-like beams are produced
predominantly through charged-pion and
kaon decays~\cite{DUNE:2020ypp}, neglecting possible NP effects in production,
we model neutrino (antineutrino) mode as an initially pure $\nu_\mu$
($\bar{\nu}_\mu$) beam. After propagation over the baseline $L_D$ to detector
$D\in\{\mathrm{ND},\mathrm{FD}\}$, the neutrino and antineutrino states are
\begin{align}
  |\nu_D(E_\nu)\rangle       & = \sum_{b=e,\mu,\tau} \left( \sum_{i=1}^3 e^{i \frac{m_{Mi}^2 L_D}{2E_\nu}} (U_{\mu i}^M)^* U_{b i}^M \right) |\nu_{b}\rangle \equiv \sum_{b} \mathcal{A}_b^D |\nu_{b}\rangle\,, \label{eq:AmpNu}                                         \\[0.2cm]
  |\bar{\nu}_D(E_\nu)\rangle & = \sum_{b=e,\mu,\tau} \left( \sum_{i=1}^3 e^{i \frac{\bar{m}_{Mi}^2 L_D}{2E_\nu}} (\bar{U}_{\mu i}^M)^* \bar{U}_{b i}^M \right) |\bar{\nu}_{b}\rangle \equiv \sum_{b} \bar{\mathcal{A}}_b^D |\bar{\nu}_{b}\rangle\,, \label{eq:AmpAntiNu}
\end{align}
where $U^M$ ($\bar{U}^M$) denotes the matter-modified
Pontecorvo--Maki--Nakagawa--Sakata matrix and $m_{Mi}$
($\bar{m}_{Mi}$) the corresponding effective mass eigenvalues for
neutrinos (antineutrinos). Due to the short baseline, oscillations are negligible at the ND, yielding $\mathcal{A}_\mu^{\mathrm{ND}}\simeq
\bar{\mathcal{A}}_\mu^{\mathrm{ND}}\simeq1$, while the other flavor amplitudes
are neglected. At the FD, we retain the full flavor superpositions in
Eqs.~\eqref{eq:AmpNu} and \eqref{eq:AmpAntiNu}.

To obtain the amplitudes $\mathcal{A}_b^D$ and
$\bar{\mathcal{A}}_b^D$, we import the DUNE-like configuration of
Ref.~\cite{DUNE:2021cuw} into the GLoBES 3.2.18
package~\cite{Huber:2004ka, Huber:2007ji}. As inputs for simulation,
we use a constant baseline matter density of
$2.848~\mathrm{g/cm^3}$~\cite{Roe:2017zdw,DUNE:2021cuw} and the best-fit
oscillation parameters from the NuFIT 6.1 global
fit~\cite{Esteban:2024eli, NuFit} summarized in
Table~\ref{tab:oscillation_parameters}.\footnote{A similar GLoBES-based DUNE
  simulation was performed in Ref.~\cite{Abbaslu:2023vqk}. Here, we adopt the
  updated NuFIT~6.1 global-fit parameters and consider both normal and inverted
mass orderings.}
The same amplitudes apply to the
Majorana case because Majorana phases cancel from standard oscillation
probabilities.

\begin{table*}[t]
  \centering
  \renewcommand{\arraystretch}{1.5}
  \setlength{\tabcolsep}{16pt}
  \caption{Three-flavor neutrino oscillation parameters for normal mass ordering (NO) and inverted mass ordering (IO) from the NuFIT 6.1 global fit~\cite{Esteban:2024eli, NuFit}, excluding atmospheric neutrino data from Super-Kamiokande and IceCube. Here, $\Delta m_{3\ell}^2 = \Delta m_{31}^2 > 0$ for NO and $\Delta m_{3\ell}^2 = \Delta m_{32}^2 < 0$ for IO.}
  \begin{tabular}{ccccccc}
    \hline\hline
    & \(\theta_{12}/^\circ\) & \(\theta_{23}/^\circ\) & \(\theta_{13}/^\circ\) & \(\delta_{\text{CP}}/^\circ\) & \(\Delta m_{12}^2 / 10^{-5} \mathrm{eV}^2\) & \(\Delta m_{3\ell}^2 / 10^{-3} \mathrm{eV}^2\) \\
    \hline
    NO& $33.76^{+0.42}_{-0.41}$ & $43.27^{+1.0}_{-0.82}$ & $8.62^{+0.11}_{-0.11}$ & $207^{+23}_{-20}$ & $7.537^{+0.094}_{-0.10}$ & $+2.521^{+0.026}_{-0.018}$ \\
    IO & $33.76^{+0.42}_{-0.41}$ & $48.15^{+0.75}_{-0.92}$ & $8.65^{+0.11}_{-0.11}$ & $283^{+24}_{-28}$ & $7.537^{+0.094}_{-0.10}$ & $-2.500^{+0.024}_{-0.023}$ \\
    \hline\hline
  \end{tabular}
  \label{tab:oscillation_parameters}
\end{table*}

Since the outgoing neutrino is not observed in NC measurements, the cross
sections at either detector are summed over all allowed final-state neutrino
flavors,
\begin{align}
  (\sigma_N)_{\nu_D}       & = \sum_{b=e,\mu,\tau} \sigma(\nu_D+N\to\nu_b+X)\,,             \\[0.2cm]
  (\sigma_N)_{\bar{\nu}_D} & = \sum_{b=e,\mu,\tau} \sigma(\bar{\nu}_D+N\to\bar{\nu}_b+X) \,.
\end{align}
The SM and NP contributions are then combined incoherently, as discussed
above, before convolution with beam flux.

The expected number of NC DIS events in a given detector and beam mode is then given by
\begin{equation}
  \label{eq:event_rate}
  \mathcal{N}_{\nu/\bar{\nu}}^{\mathrm{ND/FD}}
  = \int dE\,\phi_{\nu/\bar{\nu}}^{\mathrm{ND/FD}}(E)
  \left[(\sigma_n)_{\nu/\bar{\nu}}N_n^{\mathrm{ND/FD}}
  + (\sigma_p)_{\nu/\bar{\nu}}N_p^{\mathrm{ND/FD}}\right] \,.
\end{equation}
Here $\phi_{\nu/\bar{\nu}}^{\mathrm{ND/FD}}(E)$ denotes the time-integrated flux
at the corresponding detector. We use the DUNE flux predictions from
Ref.~\cite{DUNE_flux}, quoted in units of
$\mathrm{GeV}^{-1}\,\mathrm{m}^{-2}$ per proton on target (POT), for both the
``CP-optimized'' (the ``Nov 2017'' version) and ``$\tau$-optimized'' (the ``Jan 2021''
version) beam configurations. The exposure is taken to be
$1.1\times10^{21}$~POT/year, with 6.5 years of running in each beam mode.

\section{Sensitivity Analysis and Numerical Results}
\label{sec:sensitivity}

\subsection{Statistical Methodology}
\label{sec:methodology}

To obtain expected constraints on the Wilson coefficients entering the
low-energy effective Lagrangians in Eqs.~\eqref{eq:L6D} and \eqref{eq:M6D}, we
compare the event yields including NP contributions with the SM expectation
while accounting for signal and background systematic uncertainties. Following
Ref.~\cite{Abbaslu:2023vqk}, we implement this comparison using a pull-method
$\chi^2$ test statistic based on the total NC DIS event yields in the neutrino
and antineutrino beam modes. Although that study considers flavor-conserving NC
interactions, whereas the present work focuses on $s\leftrightarrow d$ FCNCs,
both analyses rely on the projected inclusive NC DIS yields under identical DUNE-like beam and detector specifications.
The effective operators therefore modify only the predicted NP event yields, while
the treatment of efficiencies, backgrounds, and systematic uncertainties
remains unchanged. The test statistic is given by
\begin{equation}
  \chi^2 = \min_{\xi, \omega_\nu, \omega_{\bar{\nu}}} \left[ \sum_{Y = \nu, \bar{\nu}} \left( \frac{\left[ \xi \mathcal{N}_Y - \epsilon\mathcal{N}_{Y,\mathrm{SM}} + \omega_Y \mathcal{B}_Y \right]^2}{\epsilon\mathcal{N}_{Y,\mathrm{SM}}+ \mathcal{B}_Y}  + \frac{\omega_Y^2}{\sigma_\omega^2}\right) + \frac{(\xi - \epsilon)^2}{\sigma_\epsilon^2} \right] \,,
  \label{eq:chisquare}
\end{equation}
where $Y\in\{\nu,\bar{\nu}\}$ denotes the beam mode,
$\mathcal{N}_{Y,\mathrm{SM}}$ is the SM NC DIS event count, and
$\mathcal{N}_Y=\mathcal{N}_{Y,\mathrm{SM}}+\mathcal{N}_{Y,\mathrm{NP}}$ represents the total predicted event count including the NP
contribution. The minimization is performed over the signal-rate pull $\xi$ and
the beam-mode-dependent background-rate pulls $\omega_\nu$ and
$\omega_{\bar{\nu}}$.

For the signal rate, we set the nominal detection efficiency to
$\epsilon\simeq0.90$~\cite{Coloma:2017ptb} and consider
$\sigma_\epsilon=0$ and $10\%$ as benchmark signal-rate uncertainties. For
$\sigma_\epsilon=0$,
the minimization fixes $\xi=\epsilon$, equivalent to omitting the
signal-rate pull; for $\sigma_\epsilon=10\%$, $\xi$ varies about $\epsilon$
with a Gaussian prior that accounts for flux, nuclear, and detection-efficiency
uncertainties. Similarly, we take $\sigma_\omega=0$ or $2\%$ for the
background-rate uncertainty~\cite{Abbaslu:2023vqk}. The former fixes
$\omega_\nu=\omega_{\bar\nu}=0$, whereas the latter allows the background
pulls to vary with the corresponding Gaussian prior.
The background is modeled as the sum of misidentified charged-current (CC) and
resonant NC events,
\begin{equation}
  \mathcal{B}_Y^D = \epsilon_{\mathrm{CC}}\mathcal{N}_{Y,\mathrm{CC}}^D
  + \epsilon_{\mathrm{Res}}\mathcal{N}_{Y,\mathrm{Res}}^D\,.
  \label{eq:background}
\end{equation}
The factors $\epsilon_{\mathrm{CC}}$ and $\epsilon_{\mathrm{Res}}$ are the respective
misidentification probabilities, fixed to
$10\%$~\cite{Coloma:2017ptb,Tingey:2022evd}. For each detector and beam
mode, we take
$\mathcal{N}_{Y,\mathrm{CC}}^D \simeq
(71/12)\mathcal{N}_{Y,\mathrm{SM}}^D$ and
$\mathcal{N}_{Y,\mathrm{Res}}^D \simeq
(7/12)\mathcal{N}_{Y,\mathrm{SM}}^D$~\cite{DUNE:2020ypp}.
Further details regarding the numerical inputs entering
Eq.~\eqref{eq:chisquare} can be found in Ref.~\cite{Abbaslu:2023vqk}
and references therein.

\subsection{Projected Sensitivities to FCNC Couplings}

\begin{figure*}[h]
  \centering
  \includegraphics[width=0.8\textwidth]{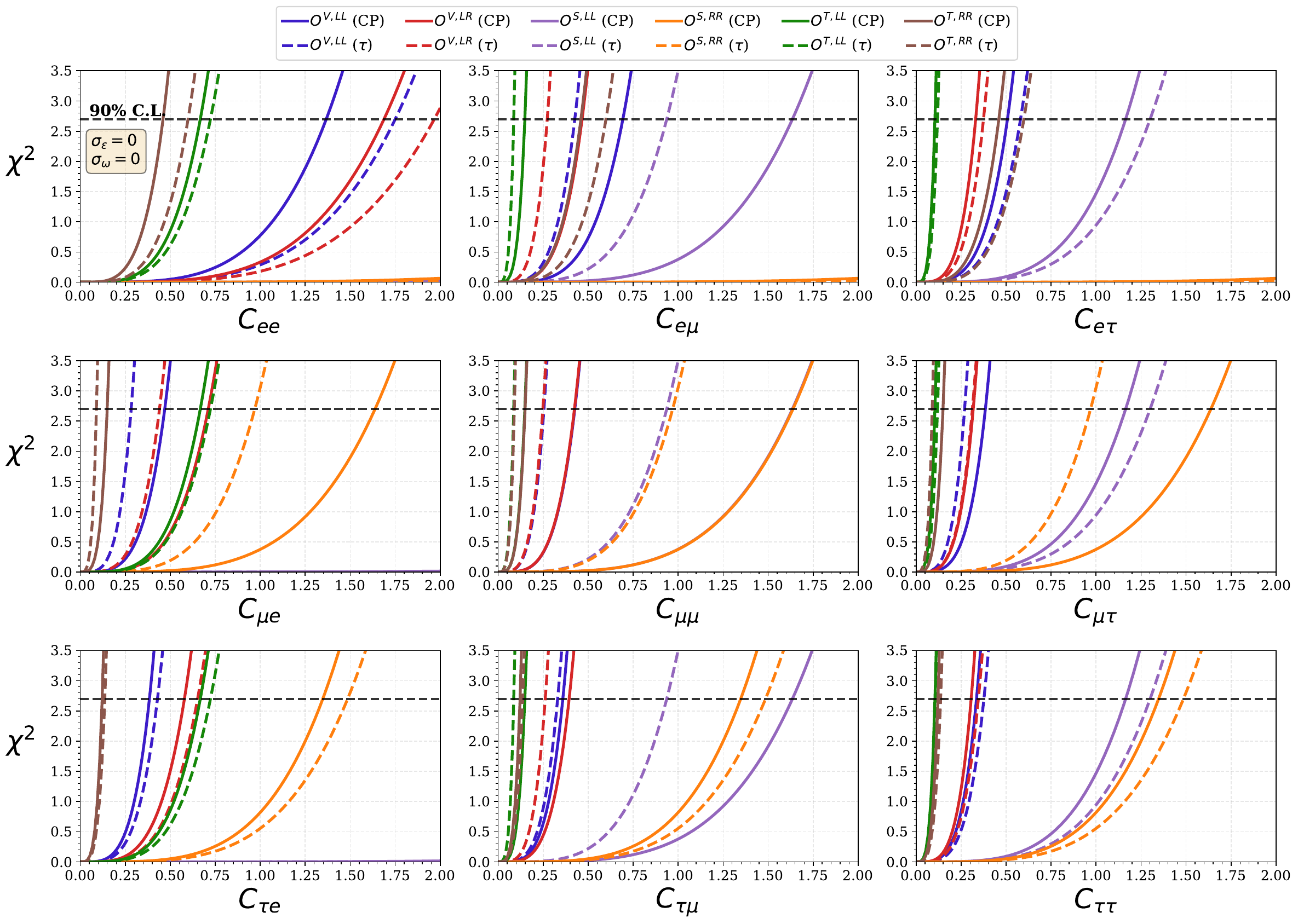}
  \par\vspace{0.2em}
  \includegraphics[width=0.8\textwidth]{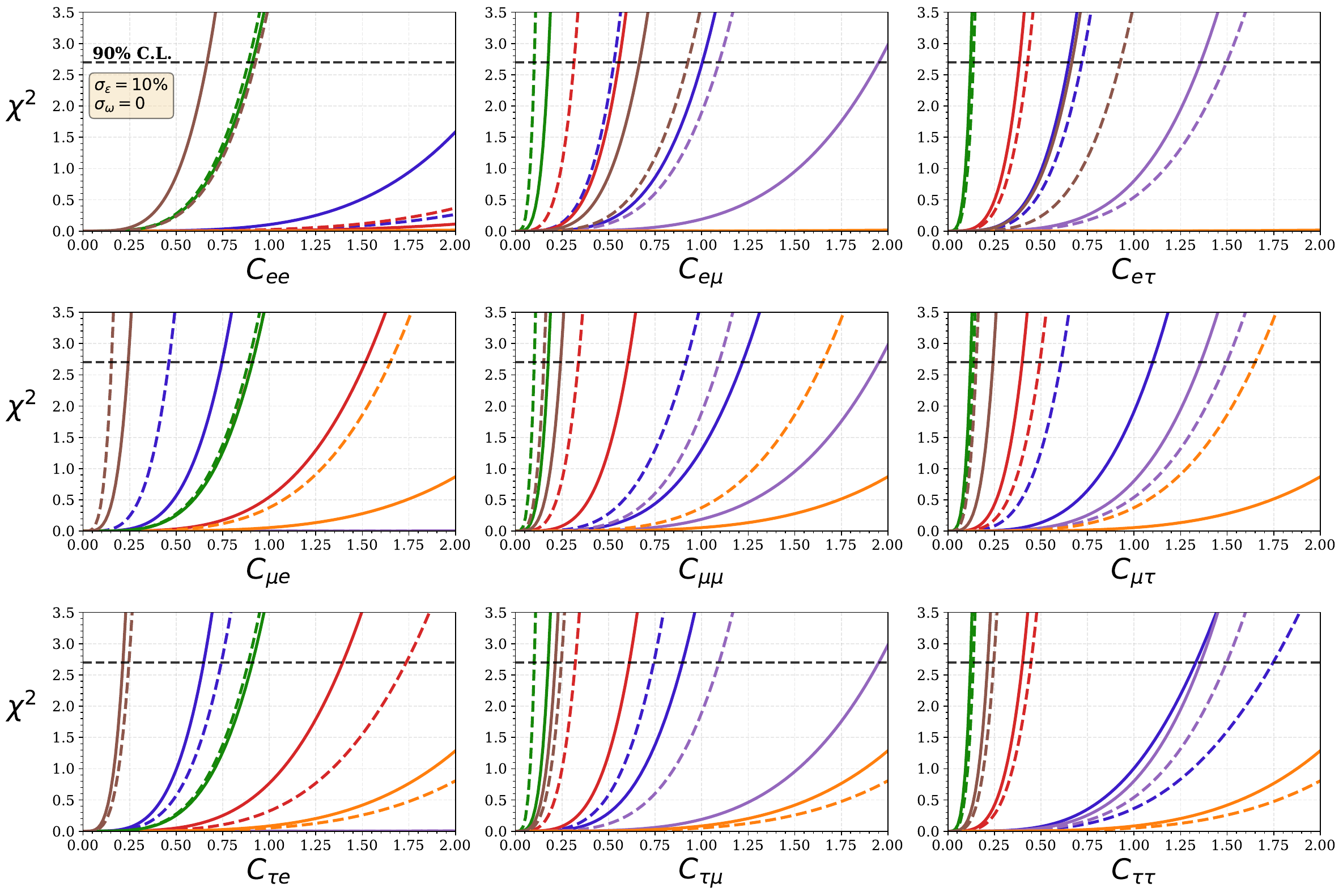}
  \caption{Projected sensitivities to the Dirac-neutrino couplings $C_{ab}^X$ [Eq.~\eqref{eq:L6D}] at the FD, assuming NO. The upper and lower panels correspond to $(\sigma_\epsilon,\sigma_\omega)=(0,0)$ and $(10\%,0)$, respectively. Colors distinguish the Lorentz and chiral structures indicated in the legend; solid and dashed curves denote the CP- and $\tau$-optimized beam configurations, respectively. The horizontal black dashed line marks $\chi^2=2.71$, corresponding to 90\% C.L. for one degree of freedom. Its intersection with each curve gives the projected upper bound. Only non-negative coefficients are displayed because the curves are symmetric around zero. Complete bounds for all three systematic scenarios and both mass orderings are given in Table~\ref{tab:Chi2_D_FD}.}
  \label{fig:Chi2_D_FD_NO}
\end{figure*}

\begin{figure*}[h]
  \centering
  \includegraphics[width=0.70\textwidth]{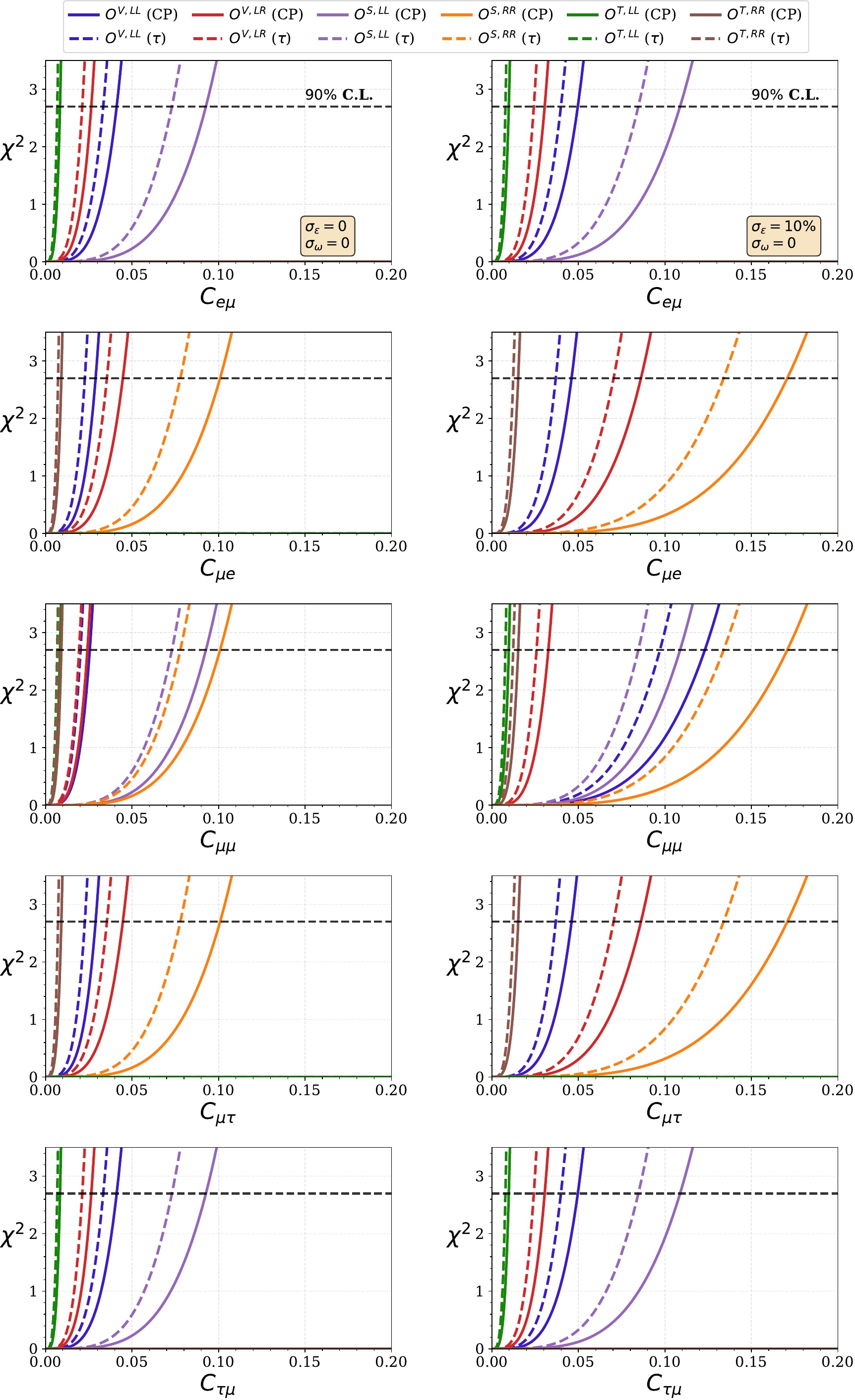}
  \caption{Dirac-neutrino sensitivities at the ND, with the same curve conventions as Fig.~\ref{fig:Chi2_D_FD_NO}. The left and right columns correspond to $(\sigma_\epsilon,\sigma_\omega)=(0,0)$ and $(10\%,0)$, respectively. Complete bounds are given in Table~\ref{tab:Chi2_D_ND}.}
  \label{fig:Chi2_D_ND}
\end{figure*}

\begin{figure*}[h]
  \centering
  \includegraphics[width=0.8\textwidth]{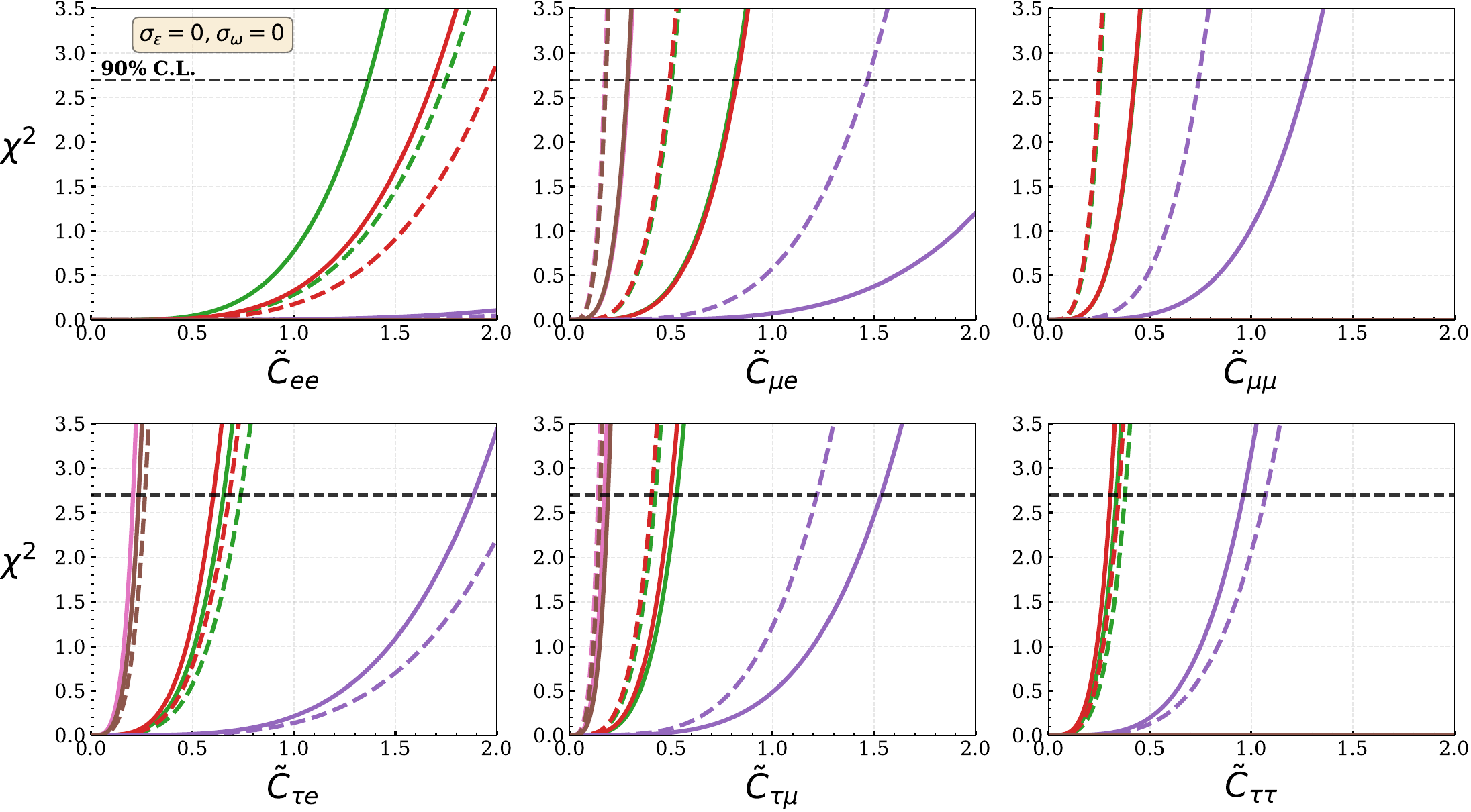}
  \par\vspace{0.5em}
  \includegraphics[width=0.8\textwidth]{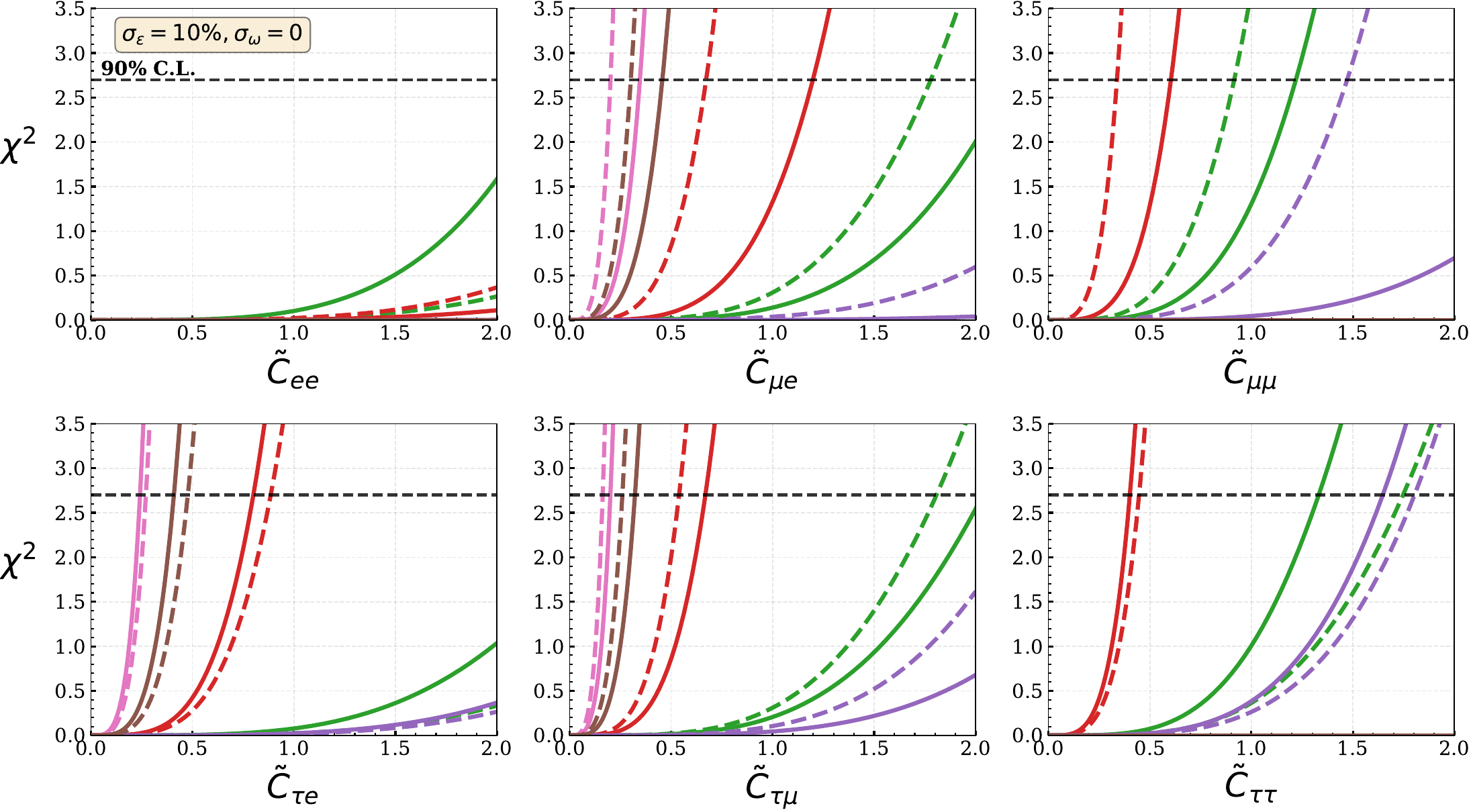}
  \caption{Projected sensitivities to the Majorana-neutrino couplings $\tilde C_{ab}^X$ [Eq.~\eqref{eq:M6D}] at the FD, assuming NO. The upper and lower panels correspond to $(\sigma_\epsilon,\sigma_\omega)=(0,0)$ and $(10\%,0)$, respectively. Curve styles and the 90\% C.L. threshold follow Fig.~\ref{fig:Chi2_D_FD_NO}. Complete bounds are given in Table~\ref{tab:Chi2_M_FD}.}
  \label{fig:Chi2_M_FD_NO}
\end{figure*}

\begin{figure*}[h]
  \centering
  \includegraphics[width=0.8\textwidth]{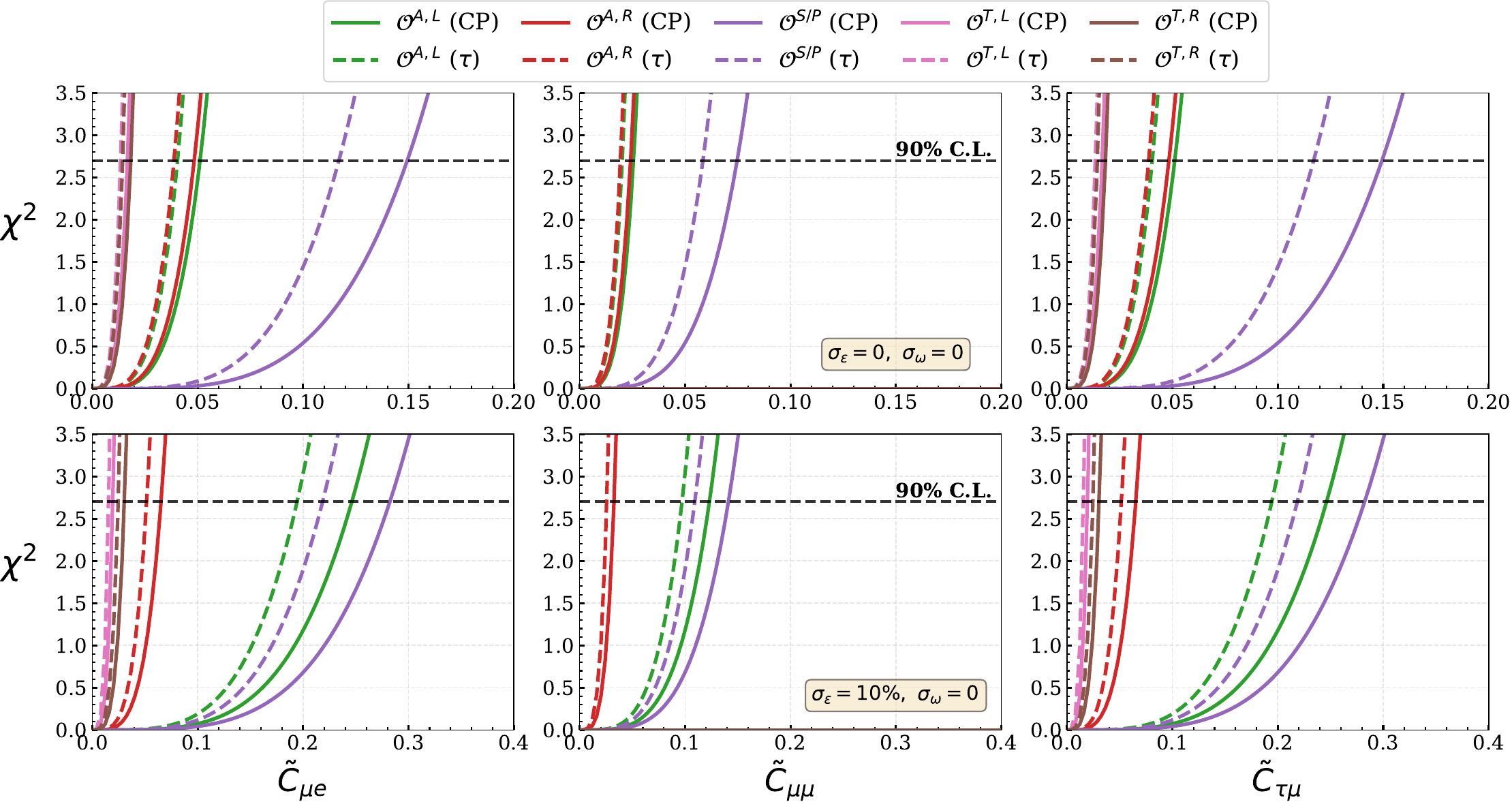}
  \caption{Majorana-neutrino sensitivities at the ND, with the same curve conventions as Fig.~\ref{fig:Chi2_M_FD_NO}. The upper and lower rows correspond to $(\sigma_\epsilon,\sigma_\omega)=(0,0)$ and $(10\%,0)$, respectively. Complete bounds are given in Table~\ref{tab:Chi2_M_ND}.}
  \label{fig:Chi2_M_ND}
\end{figure*}

We evaluate the sensitivity to each Wilson coefficient individually, setting
all other coefficients to zero. Representative
one-dimensional $\chi^2$ curves are shown in
Figs.~\ref{fig:Chi2_D_FD_NO}--\ref{fig:Chi2_M_ND}, and the complete upper
limits are collected in
Tables~\ref{tab:Chi2_D_FD}--\ref{tab:Chi2_M_ND}. To keep the figures concise,
we plot only the $(\sigma_\epsilon,\sigma_\omega)=(0,0)$ and $(10\%,0)$
benchmark scenarios, whereas the tables report the full set of results,
including the $(10\%,2\%)$ scenario. For the FD, we present figures for the
NO only and tabulate results for both the NO and IO. For the ND, by contrast,
oscillation and matter effects are negligible over the short
baseline, so NO and IO yield identical results. Across all panels, line colors
distinguish the Lorentz and chiral structures, solid (dashed) curves denote the
CP- ($\tau$-)optimized flux, and the horizontal dashed line marks the 90\%
C.L. threshold ($\chi^2=2.71$ for one degree of freedom). Since the SM contributions to our FCNC processes are neglected,
these $\chi^2$ curves are symmetric about zero, and only non-negative values of the
Wilson coefficients are shown.

\subsubsection{Dirac-Neutrino Sensitivities}

Several degeneracies arise among the projected bounds for different flavor
combinations, as shown in Tables~\ref{tab:Chi2_D_FD} and
\ref{tab:Chi2_D_ND}. At the FD, the limits on the scalar
and tensor coefficients $C_{ab}^{X,LL}$ ($X\in\{S,T\}$) depend only on the
second flavor index $b$, whereas those on $C_{ab}^{X,RR}$ depend solely on the
first index $a$. For example, identical limits are obtained for
$C_{ee}^{S,LL}$, $C_{\mu e}^{S,LL}$, and $C_{\tau e}^{S,LL}$, and likewise for
$C_{ee}^{S,RR}$, $C_{e\mu}^{S,RR}$, and $C_{e\tau}^{S,RR}$. The corresponding
tensor bounds follow the same flavor-index pattern, which also persists
at the ND. The nearly pure muon-flavor composition of the ND flux produces the
additional relations that identical vector limits are obtained for the pairs
$C_{e\mu}^{V,L\kappa}$ and $C_{\tau\mu}^{V,L\kappa}$, as well as
$C_{\mu e}^{V,L\kappa}$ and $C_{\mu\tau}^{V,L\kappa}$
($\kappa\in\{L,R\}$). These degeneracies arise from the chiral projectors in
the effective operators, which determine which flavor index labels the
incident state.

In general, the tensor coefficients are the most tightly constrained and the
scalar coefficients the least, with the vector limits falling in between.
Exceptions occur, however. For example, at the
FD with the CP-optimized flux, assuming NO and
$(\sigma_\epsilon,\sigma_\omega)=(0,0)$, the upper limits on
$C_{\mu e}^{V,LL}$ and $C_{\tau e}^{V,LL}$ are $0.47$ and $0.38$,
respectively, both stronger than the corresponding tensor limit of $0.67$ for
each coefficient. Despite these exceptions, the
overall sensitivity hierarchy is reversed relative to that in the rare kaon decay
$K^+\to\pi^+\nu\bar\nu$, where scalar operators are the most strongly
constrained and tensor operators the least~\cite{Gorbahn:2023juq}.

Among all Dirac-neutrino Wilson coefficients considered, the most stringent
bounds at both the ND and the FD are achieved with the $\tau$-optimized flux in
the $(\sigma_\epsilon,\sigma_\omega)=(0,0)$ scenario. The leading constraints on the tensor, vector, and scalar coefficients at the ND and FD are, respectively,
\begin{align}
  \text{ND:} \quad & |C_{a\mu}^{T,LL}| \le 0.0068\,, \qquad |C_{\mu\mu}^{V,LR}| \le 0.019\,, \qquad |C_{a\mu}^{S,LL}| \le 0.073\,, \\[0.2cm]
  \text{FD:} \quad & |C_{a\mu}^{T,LL}| \le 0.087\,, \qquad |C_{\mu\mu}^{V,L\kappa}| \le 0.25\,, \qquad |C_{a\mu}^{S,LL}| \le 0.94\,,
\end{align}
where $a \in \{e, \mu, \tau\}$.
As expected, the ND yields substantially tighter bounds than the FD owing to
its larger event statistics. Nevertheless, these projected sensitivities remain
less stringent than indirect limits from
$K^+\to\pi^+\nu\bar\nu$ decay, where the constraints on tensor, vector, and scalar
coefficients reach $\mathcal{O}(10^{-4})$, $\mathcal{O}(10^{-5})$, and
$\mathcal{O}(10^{-6})$, respectively~\cite{Gorbahn:2023juq}. We note, however,
that the limits from $K^+\to\pi^+\nu\bar\nu$ apply to Wilson coefficients
defined in the neutrino mass basis, whereas the sensitivities derived
here directly probe Wilson coefficients defined in the neutrino flavor basis.

Including signal- and background-rate uncertainties generally weakens the
projected sensitivities. At the FD, introducing the signal-rate
uncertainty in the $(\sigma_\epsilon,\sigma_\omega)=(10\%,0)$ scenario weakens
most scalar and tensor limits by several tens of percent, whereas the vector
bounds are weakened substantially. For example, while
left-handed scalar and tensor constraints typically worsen by less than
$40\%$, the limit on $C_{\tau\tau}^{V,LL}$ loosens by up to $400\%$.
Further including a background-rate uncertainty of $\sigma_\omega=2\%$
weakens the FD bounds by only an additional few tens of percent.
Consequently, the cumulative weakening relative to the idealized $(0,0)$ case
ranges from $36\%$ to $170\%$ for scalar and tensor bounds, while the
maximum sensitivity loss among vector coefficients remains at $400\%$.

In contrast to the FD, the ND limits are considerably more sensitive to the
background-rate uncertainty. For $(\sigma_\epsilon,\sigma_\omega)=(10\%,0)$,
the bounds on the tightly constrained left-handed $e\mu$ and $\tau\mu$
coefficients increase by only $14\%\text{--}22\%$, whereas some vector bounds
increase more substantially. For example, with the CP-optimized flux, the
upper limit on $C_{\mu\mu}^{V,LL}$ increases from $0.026$ to $0.12$, or by
approximately $360\%$. Once $\sigma_\omega=2\%$ is incorporated,
all constrained ND limits weaken dramatically, worsening by factors of
roughly $13\text{--}49$ relative to the idealized $(0,0)$ case.

We next examine how the assumed neutrino mass ordering affects the projected
sensitivities.
At the ND, flavor oscillation effects are negligible owing to the short
baseline, and the NO and IO bounds
therefore coincide. At
the FD, the mass-ordering dependence is mild for most coefficients, typically
remaining at or below the few-percent level. Substantial differences arise
primarily in Wilson coefficients involving electron flavor. For example, under
the CP-optimized flux with $(\sigma_\epsilon,\sigma_\omega)=(10\%,0)$, the
$C_{ee}^{V,LL}$ and $C_{ee}^{V,LR}$ limits differ between NO and IO by $78\%$
and $65\%$, respectively, where the relative difference is defined as
$R \equiv (\text{NO}-\text{IO})/\text{NO}$. Large differences also appear
for several off-diagonal couplings involving the electron flavor, reaching
$48\%$ for $C_{\mu e}^{V,LR}$ and approximately $44\%$ for $C_{\mu e}^{S,LL}$.
By contrast, coefficients involving exclusively muon and tau flavors are
practically insensitive to the mass ordering.

Finally, we consider the impact of uncertainties in the neutrino oscillation
parameters listed in Table~\ref{tab:oscillation_parameters}. The mixing angles
and mass-squared differences have relative uncertainties of $\mathcal{O}(1\%)$,
whereas the CP-violating phase $\delta_{\text{CP}}$ has a relative uncertainty
of $\mathcal{O}(10\%)$. We therefore focus on its effect on the projected FD
sensitivity. For a fixed mass ordering and
experimental configuration, we vary $\delta_{\text{CP}}$ within its $1\sigma$
interval and quantify the resulting fractional shift as
\begin{equation}\label{eq:ratio}
  r\equiv
  \frac{C_{90}(\delta_{\mathrm{CP}})-
  C_{90}(\delta_{\mathrm{CP}}^{0})}
  {C_{90}(\delta_{\mathrm{CP}}^{0})}\,,
\end{equation}
where $\delta_{\mathrm{CP}}^{0}$ is the central value listed in
Table~\ref{tab:oscillation_parameters}, and $C_{90}$ denotes the 90\% C.L.
upper bound on a given Wilson coefficient. For illustration, we focus on NO and
find that the $C_{\mu\mu}$ coefficients remain
essentially insensitive, with $|r|$ of order $10^{-3}$, while
$C_{\mu\tau}$, $C_{\tau\mu}$, and $C_{\tau\tau}$ satisfy $|r|<1\%$ for
$(\sigma_\epsilon,\sigma_\omega)=(0,0)$ and reach at most a few percent when
rate uncertainties are included. The largest variations occur among
$C_{ee}$, $C_{e\mu}$, $C_{e\tau}$, $C_{\mu e}$, and $C_{\tau e}$, for which the maximal
$|r|$ is approximately $10\%\text{--}20\%$ in the idealized $(0,0)$ case and
can reach about $70\%$ ($90\%$) for $C_{ee}^{V,LL}$ with the CP-optimized
($\tau$-optimized) flux when $(\sigma_\epsilon,\sigma_\omega)=(10\%,0)$.

\subsubsection{Majorana-Neutrino Sensitivities}

As in the Dirac case, the projected Majorana bounds display several
degeneracies among different Lorentz and flavor structures, as shown in
Tables~\ref{tab:Chi2_M_FD} and \ref{tab:Chi2_M_ND}.
First, as dictated by the Majorana symmetry relations in
Eq.~\eqref{eq:symmetry}, the flavor-diagonal vector and tensor coefficients
identically vanish. For each off-diagonal flavor combination ($a\neq b$),
identical limits are obtained for $\tilde C_{ab}^{V,\kappa}$ and
$\tilde C_{ab}^{A,\kappa}$ at fixed quark chirality. The limits on
$\tilde C_{ab}^{S,L}$, $\tilde C_{ab}^{S,R}$, $\tilde C_{ab}^{P,L}$, and
$\tilde C_{ab}^{P,R}$ are likewise identical. At the ND, the nearly pure
$\nu_\mu$ beam additionally yields identical limits for
$\tilde C_{\mu e}^{X,\kappa}$ and $\tilde C_{\tau\mu}^{X,\kappa}$, where
$X\in\{V,A,S,P,T\}$. These degeneracies arise because neutrinos are treated as massless in the
scattering cross sections and only one Wilson coefficient is varied at a time.

Beyond these degeneracies, the overall sensitivity hierarchy mirrors
that of the Dirac case: tensor interactions are generally subject to the most
stringent constraints, scalar and pseudoscalar interactions to the weakest,
while vector and axial-vector limits fall in between. To quantify the best
attainable reach, we first consider the idealized
$(\sigma_\epsilon,\sigma_\omega)=(0,0)$ benchmark, for which the strongest
Majorana bounds are obtained with the $\tau$-optimized flux. At the ND, they
are
\begin{equation}
  \begin{aligned}
    |\tilde C_{\mu e}^{T,\kappa}| \le 0.014\,, \qquad
    |\tilde C_{\tau\mu}^{T,\kappa}| \le 0.014\,, \qquad
    |\tilde C_{\mu\mu}^{A,R}| \le 0.019\,, \qquad
    |\tilde C_{\mu\mu}^{S,\kappa}| \le 0.059\,, \qquad
    |\tilde C_{\mu\mu}^{P,\kappa}| \le 0.059\,.
  \end{aligned}
\end{equation}
The strongest FD bounds are
\begin{equation}
  \begin{aligned}
    |\tilde C_{\tau\mu}^{T,L}| &\le 0.14\,, \qquad
    |\tilde C_{\mu\mu}^{A,\kappa}| \le 0.25\,, \qquad
    |\tilde C_{\mu\mu}^{S,\kappa}| \le 0.74\,, \qquad
    |\tilde C_{\mu\mu}^{P,\kappa}| \le 0.74\,.
  \end{aligned}
\end{equation}
These benchmark results show that the ND improves the best Majorana
sensitivities by approximately one order of magnitude relative to the FD.
Nevertheless, the resulting scattering sensitivities remain considerably
weaker than indirect limits from
$K^+\to\pi^+\nu\bar\nu$~\cite{Gorbahn:2023juq}, which reach $\mathcal{O}(10^{-4})$,
$\mathcal{O}(10^{-5})$, and $\mathcal{O}(10^{-6})$ for tensor, vector/axial-vector,
and scalar/pseudoscalar coefficients, respectively. As noted in the Dirac
discussion, these decay limits apply to mass-basis coefficients, whereas our
results constrain the flavor basis.

Having established the idealized reach, we next assess its robustness against
signal- and background-rate uncertainties. At the FD, introducing
$\sigma_\epsilon=10\%$ substantially
relaxes the vector, axial-vector, and scalar bounds. For instance, under the
$\tau$-optimized flux, the $\tilde C_{\tau e}^{V,L}$ and $\tilde C_{ee}^{S,L}$
limits loosen by approximately $390\%$ and $360\%$, respectively. In contrast,
the $\tilde C_{\tau\mu}^{T,L}$ bound weakens by only about $14\%$, while its
right-handed counterpart worsens by about $73\%$. Incorporating
$\sigma_\omega=2\%$ yields only a modest further loss of sensitivity at the FD.
At the ND, however, the background-rate uncertainty is again the dominant
systematic: once $\sigma_\omega=2\%$ is included, the constrained Majorana
bounds weaken by factors of roughly $15\text{--}49$ relative to the idealized
$(0,0)$ case.

Beyond these detector systematics, the projected Majorana sensitivities at the
FD can depend on the assumed neutrino mass ordering as well. For most Wilson
coefficients, the bounds differ by only a few percent between NO and IO,
following the same general pattern as in the Dirac case.
Larger variations are concentrated among the $\tilde C_{ee}$ coefficients. In the
$(\sigma_\epsilon,\sigma_\omega)=(10\%,0)$ case, the relative difference
$R$ reaches $78\%$ for $\tilde C_{ee}^{A,L}$ and $65\%$ for
$\tilde C_{ee}^{A,R}$ under the CP-optimized flux, and $69\%$ for
$\tilde C_{ee}^{S,L}$ under the $\tau$-optimized flux. For all remaining
couplings, the dependence on the mass ordering does not exceed $14\%$.

We finally examine the dependence on $\delta_{\text{CP}}$ for NO using the
ratio $r$ defined in Eq.~\eqref{eq:ratio}. The $\tilde C_{\mu\mu}$
coefficients are the least affected, with $|r|$ remaining of order $10^{-3}$
in all three systematic scenarios. Across these scenarios, the
$\tilde C_{\tau\mu}$, $\tilde C_{\tau\tau}$, and $\tilde C_{\mu e}$
coefficients show only mild variations, with maximal $|r|$ values of several
percent, while those for
$\tilde C_{\tau e}$ remain below $0.1\%$. The strongest
dependence occurs among the $\tilde C_{ee}$ coefficients. In particular, for
$\tilde C_{ee}^{A,L}$, $|r|$ is of order $10\%$ in the idealized case and
reaches about $70\%$ ($90\%$) with the CP-optimized ($\tau$-optimized) flux
when $(\sigma_\epsilon,\sigma_\omega)=(10\%,0)$. After including
$\sigma_\omega=2\%$, it remains about $70\%$ for either flux.

\begin{table*}[t]
  \renewcommand{\arraystretch}{1.3}

  \caption{Projected 90\% C.L. upper bounds on the Dirac-neutrino Wilson coefficients $C_{ab}^X$ [Eq.~\eqref{eq:L6D}] at the FD of a DUNE-like experiment, assuming 6.5 years of exposure in each of the $\nu$ and $\bar\nu$ modes and varying one coefficient at a time. Results are shown for normal (NO) and inverted (IO) mass ordering, with $R=(\text{NO}-\text{IO})/\text{NO}$. Since the confidence intervals are symmetric about zero, only their positive limits are reported. The bounds on $C_{ab}^{S,LR}$ and $C_{ab}^{S,RL}$ coincide with those on $C_{ab}^{S,LL}$ and $C_{ab}^{S,RR}$, respectively, and are therefore omitted. All values are rounded to two significant figures.}
  \label{tab:Chi2_D_FD}

  \begin{ruledtabular}
    \begin{tabular}{ll ccc ccc ccc}
      & & \multicolumn{3}{c}{$(\sigma_\epsilon, \sigma_\omega) = (0, 0)$} & \multicolumn{3}{c}{$(\sigma_\epsilon, \sigma_\omega) = (10\%, 0)$} & \multicolumn{3}{c}{$(\sigma_\epsilon, \sigma_\omega) = (10\%, 2\%)$} \\
      \cmidrule(lr){3-5} \cmidrule(lr){6-8} \cmidrule(lr){9-11}
      Coupling & Mode & NO & IO & $R$ (\%) & NO & IO & $R$ (\%) & NO & IO & $R$ (\%) \\
      \hline

      $C_{ee}^{V,LL}$ & CP  & 1.4 & 1.5 & -9.2 & 2.3 & 4.1 & -78 & 2.8 & 4.7 & -68 \\
      & $\tau$ & 1.8 & 1.7 & +1.5 & 3.6 & 5.9 & -63 & 5.0 & 7.3 & -45 \\
      $C_{ee}^{V,LR}$ & CP  & 1.7 & 1.3 & +24 & 4.6 & 1.6 & +65 & 5.3 & 2.0 & +63 \\
      & $\tau$ & 2.0 & 1.5 & +22 & 3.3 & 1.9 & +42 & 4.6 & 2.7 & +41 \\
      $C_{ee}^{S,LL}$ & CP  & 7.3 & 4.8 & +34 & 10 & 5.5 & +44 & 12 & 6.8 & +44 \\
      & $\tau$ & 7.8 & 5.7 & +27 & 9.6 & 6.5 & +32 & 14 & 9.2 & +32 \\
      $C_{ee}^{S,RR}$ & CP  & 5.1 & 6.1 & -20 & 7.4 & 12 & -61 & 9.1 & 15 & -60 \\
      & $\tau$ & 6.5 & 6.8 & -5.0 & 10 & 13 & -29 & 14 & 18 & -28 \\
      $C_{ee}^{T,LL}$ & CP  & 0.67 & 0.44 & +35 & 0.91 & 0.50 & +45 & 1.1 & 0.62 & +45 \\
      & $\tau$ & 0.72 & 0.53 & +27 & 0.89 & 0.60 & +32 & 1.3 & 0.85 & +32 \\
      $C_{ee}^{T,RR}$ & CP  & 0.46 & 0.56 & -21 & 0.67 & 1.1 & -62 & 0.82 & 1.3 & -60 \\
      & $\tau$ & 0.60 & 0.63 & -5.7 & 0.93 & 1.2 & -29 & 1.3 & 1.7 & -29 \\
      \hline

      $C_{e\mu}^{V,LL}$ & CP  & 0.69 & 0.70 & -0.87 & 1.0 & 0.95 & +5.6 & 1.2 & 1.2 & +5.5 \\
      & $\tau$ & 0.43 & 0.43 & +0.00 & 0.53 & 0.53 & +0.43 & 0.75 & 0.74 & +0.43 \\
      $C_{e\mu}^{V,LR}$ & CP  & 0.47 & 0.47 & -0.090 & 0.56 & 0.55 & +0.63 & 0.69 & 0.68 & +0.63 \\
      & $\tau$ & 0.27 & 0.27 & -0.070 & 0.31 & 0.31 & -0.030 & 0.44 & 0.44 & -0.050 \\
      $C_{e\mu}^{S,LL}$ & CP  & 1.6 & 1.6 & -0.40 & 2.0 & 2.0 & -0.50 & 2.4 & 2.4 & -0.50 \\
      & $\tau$ & 0.94 & 0.94 & -0.17 & 1.1 & 1.1 & -0.17 & 1.5 & 1.5 & -0.18 \\
      $C_{e\mu}^{S,RR}$ & CP  & 5.1 & 6.1 & -20 & 7.4 & 12 & -61 & 9.1 & 15 & -60 \\
      & $\tau$ & 6.5 & 6.8 & -5.0 & 10 & 13 & -29 & 14 & 18 & -28 \\
      $C_{e\mu}^{T,LL}$ & CP  & 0.15 & 0.15 & -0.34 & 0.18 & 0.18 & -0.40 & 0.22 & 0.22 & -0.41 \\
      & $\tau$ & 0.087 & 0.087 & -0.11 & 0.10 & 0.10 & -0.20 & 0.14 & 0.14 & -0.14 \\
      $C_{e\mu}^{T,RR}$ & CP  & 0.46 & 0.56 & -21 & 0.67 & 1.1 & -62 & 0.82 & 1.3 & -60 \\
      & $\tau$ & 0.60 & 0.63 & -5.7 & 0.93 & 1.2 & -29 & 1.3 & 1.7 & -29 \\
      \hline

      $C_{e\tau}^{V,LL}$ & CP  & 0.51 & 0.51 & -1.4 & 0.65 & 0.65 & +0.59 & 0.80 & 0.80 & +0.55 \\
      & $\tau$ & 0.58 & 0.58 & -0.57 & 0.72 & 0.72 & -0.18 & 1.0 & 1.0 & -0.19 \\
      $C_{e\tau}^{V,LR}$ & CP  & 0.33 & 0.34 & -1.5 & 0.38 & 0.39 & -1.4 & 0.47 & 0.48 & -1.4 \\
      & $\tau$ & 0.37 & 0.38 & -0.83 & 0.43 & 0.43 & -0.89 & 0.60 & 0.61 & -0.88 \\
      $C_{e\tau}^{S,LL}$ & CP  & 1.2 & 1.2 & -1.5 & 1.4 & 1.4 & -1.9 & 1.7 & 1.7 & -1.9 \\
      & $\tau$ & 1.3 & 1.3 & -0.90 & 1.5 & 1.5 & -1.1 & 2.1 & 2.1 & -1.1 \\
      $C_{e\tau}^{S,RR}$ & CP  & 5.1 & 6.1 & -20 & 7.4 & 12 & -61 & 9.1 & 15 & -60 \\
      & $\tau$ & 6.5 & 6.8 & -5.0 & 10 & 13 & -29 & 14 & 18 & -28 \\
      $C_{e\tau}^{T,LL}$ & CP  & 0.11 & 0.11 & -1.5 & 0.12 & 0.12 & -1.9 & 0.15 & 0.15 & -1.9 \\
      & $\tau$ & 0.12 & 0.12 & -0.92 & 0.14 & 0.14 & -1.2 & 0.19 & 0.20 & -1.1 \\
      $C_{e\tau}^{T,RR}$ & CP  & 0.46 & 0.56 & -21 & 0.67 & 1.1 & -62 & 0.82 & 1.3 & -60 \\
      & $\tau$ & 0.60 & 0.63 & -5.7 & 0.93 & 1.2 & -29 & 1.3 & 1.7 & -29 \\
    \end{tabular}
  \end{ruledtabular}
\end{table*}

\begin{table*}[t]
  \addtocounter{table}{-1}

  \renewcommand{\arraystretch}{1.3}

  \caption{Sensitivity bounds on couplings $C_{\alpha\beta}^{X}$ at the FD (Continued).}

  \begin{ruledtabular}
    \begin{tabular}{ll ccc ccc ccc}
      & & \multicolumn{3}{c}{$(\sigma_\epsilon, \sigma_\omega) = (0, 0)$} & \multicolumn{3}{c}{$(\sigma_\epsilon, \sigma_\omega) = (10\%, 0)$} & \multicolumn{3}{c}{$(\sigma_\epsilon, \sigma_\omega) = (10\%, 2\%)$} \\
      \cmidrule(lr){3-5} \cmidrule(lr){6-8} \cmidrule(lr){9-11}
      Coupling  & Mode & NO & IO & $R$ (\%) & NO & IO & $R$ (\%) & NO & IO & $R$ (\%) \\
      \hline

      $C_{\mu e}^{V,LL}$ & CP  & 0.47 & 0.47 & +0.13 & 0.75 & 0.77 & -3.2 & 0.92 & 0.95 & -3.1 \\
      & $\tau$ & 0.28 & 0.28 & -0.11 & 0.46 & 0.46 & -0.80 & 0.65 & 0.65 & -0.79 \\
      $C_{\mu e}^{V,LR}$ & CP  & 0.71 & 0.69 & +2.7 & 1.5 & 2.2 & -48 & 1.8 & 2.5 & -36 \\
      & $\tau$ & 0.44 & 0.44 & +0.39 & 0.91 & 0.97 & -7.0 & 1.3 & 1.3 & -6.4 \\
      $C_{\mu e}^{S,LL}$ & CP  & 7.3 & 4.8 & +34 & 10 & 5.5 & +44 & 12 & 6.8 & +44 \\
      & $\tau$ & 7.8 & 5.7 & +27 & 9.6 & 6.5 & +32 & 14 & 9.2 & +32 \\
      $C_{\mu e}^{S,RR}$ & CP  & 1.6 & 1.6 & +0.020 & 2.7 & 2.7 & +0.21 & 3.3 & 3.3 & +0.20 \\
      & $\tau$ & 0.97 & 0.97 & -0.14 & 1.6 & 1.7 & -0.13 & 2.3 & 2.3 & -0.13 \\
      $C_{\mu e}^{T,LL}$ & CP  & 0.67 & 0.44 & +35 & 0.91 & 0.50 & +45 & 1.1 & 0.62 & +45 \\
      & $\tau$ & 0.72 & 0.53 & +27 & 0.89 & 0.60 & +32 & 1.3 & 0.85 & +32 \\
      $C_{\mu e}^{T,RR}$ & CP  & 0.15 & 0.15 & +0.070 & 0.24 & 0.24 & +0.25 & 0.30 & 0.30 & +0.27 \\
      & $\tau$ & 0.090 & 0.090 & -0.11 & 0.15 & 0.15 & -0.13 & 0.21 & 0.21 & -0.14 \\
      \hline

      $C_{\mu\mu}^{V,LL}$ & CP  & 0.43 & 0.43 & +0.00 & 1.2 & 1.2 & +1.1 & 1.4 & 1.4 & +0.85 \\
      & $\tau$ & 0.25 & 0.25 & -0.12 & 0.92 & 0.92 & -0.030 & 1.1 & 1.1 & -0.090 \\
      $C_{\mu\mu}^{V,LR}$ & CP  & 0.42 & 0.43 & -0.26 & 0.60 & 0.61 & -0.56 & 0.74 & 0.75 & -0.55 \\
      & $\tau$ & 0.25 & 0.25 & -0.12 & 0.34 & 0.34 & -0.18 & 0.48 & 0.48 & -0.17 \\
      $C_{\mu\mu}^{S,LL}$ & CP  & 1.6 & 1.6 & -0.40 & 2.0 & 2.0 & -0.50 & 2.4 & 2.4 & -0.50 \\
      & $\tau$ & 0.94 & 0.94 & -0.17 & 1.1 & 1.1 & -0.17 & 1.5 & 1.5 & -0.18 \\
      $C_{\mu\mu}^{S,RR}$ & CP  & 1.6 & 1.6 & +0.020 & 2.7 & 2.7 & +0.21 & 3.3 & 3.3 & +0.20 \\
      & $\tau$ & 0.97 & 0.97 & -0.14 & 1.6 & 1.7 & -0.13 & 2.3 & 2.3 & -0.13 \\
      $C_{\mu\mu}^{T,LL}$ & CP  & 0.15 & 0.15 & -0.34 & 0.18 & 0.18 & -0.40 & 0.22 & 0.22 & -0.41 \\
      & $\tau$ & 0.087 & 0.087 & -0.11 & 0.10 & 0.10 & -0.20 & 0.14 & 0.14 & -0.14 \\
      $C_{\mu\mu}^{T,RR}$ & CP  & 0.15 & 0.15 & +0.070 & 0.24 & 0.24 & +0.25 & 0.30 & 0.30 & +0.27 \\
      & $\tau$ & 0.090 & 0.090 & -0.11 & 0.15 & 0.15 & -0.13 & 0.21 & 0.21 & -0.14 \\
      \hline

      $C_{\mu\tau}^{V,LL}$ & CP  & 0.39 & 0.39 & -0.44 & 1.1 & 1.2 & -6.3 & 1.3 & 1.3 & -4.6 \\
      & $\tau$ & 0.27 & 0.27 & -0.19 & 0.61 & 0.60 & +0.67 & 0.84 & 0.84 & +0.61 \\
      $C_{\mu\tau}^{V,LR}$ & CP  & 0.32 & 0.32 & -1.4 & 0.40 & 0.41 & -2.1 & 0.49 & 0.50 & -2.1 \\
      & $\tau$ & 0.31 & 0.31 & -0.58 & 0.49 & 0.50 & -1.4 & 0.69 & 0.70 & -1.4 \\
      $C_{\mu\tau}^{S,LL}$ & CP  & 1.2 & 1.2 & -1.5 & 1.4 & 1.4 & -1.9 & 1.7 & 1.7 & -1.9 \\
      & $\tau$ & 1.3 & 1.3 & -0.90 & 1.5 & 1.5 & -1.1 & 2.1 & 2.1 & -1.1 \\
      $C_{\mu\tau}^{S,RR}$ & CP  & 1.6 & 1.6 & +0.020 & 2.7 & 2.7 & +0.21 & 3.3 & 3.3 & +0.20 \\
      & $\tau$ & 0.97 & 0.97 & -0.14 & 1.6 & 1.7 & -0.13 & 2.3 & 2.3 & -0.13 \\
      $C_{\mu\tau}^{T,LL}$ & CP  & 0.11 & 0.11 & -1.5 & 0.12 & 0.12 & -1.9 & 0.15 & 0.15 & -1.9 \\
      & $\tau$ & 0.12 & 0.12 & -0.92 & 0.14 & 0.14 & -1.2 & 0.19 & 0.20 & -1.1 \\
      $C_{\mu\tau}^{T,RR}$ & CP  & 0.15 & 0.15 & +0.070 & 0.24 & 0.24 & +0.25 & 0.30 & 0.30 & +0.27 \\
      & $\tau$ & 0.090 & 0.090 & -0.11 & 0.15 & 0.15 & -0.13 & 0.21 & 0.21 & -0.14 \\
    \end{tabular}
  \end{ruledtabular}
\end{table*}

\begin{table*}[t]
  \addtocounter{table}{-1}

  \renewcommand{\arraystretch}{1.3}

  \caption{Sensitivity bounds on couplings $C_{\alpha\beta}^{X}$ at the FD (Continued).}

  \begin{ruledtabular}
    \begin{tabular}{ll ccc ccc ccc}
      & & \multicolumn{3}{c}{$(\sigma_\epsilon, \sigma_\omega) = (0, 0)$} & \multicolumn{3}{c}{$(\sigma_\epsilon, \sigma_\omega) = (10\%, 0)$} & \multicolumn{3}{c}{$(\sigma_\epsilon, \sigma_\omega) = (10\%, 2\%)$} \\
      \cmidrule(lr){3-5} \cmidrule(lr){6-8} \cmidrule(lr){9-11}
      Coupling & Mode & NO & IO & $R$ (\%) & NO & IO & $R$ (\%) & NO & IO & $R$ (\%) \\
      \hline

      $C_{\tau e}^{V,LL}$ & CP  & 0.38 & 0.38 & +1.1 & 0.65 & 0.65 & +0.17 & 0.80 & 0.79 & +0.19 \\
      & $\tau$ & 0.43 & 0.43 & +0.68 & 0.74 & 0.74 & -0.19 & 1.0 & 1.0 & -0.18 \\
      $C_{\tau e}^{V,LR}$ & CP  & 0.58 & 0.56 & +2.6 & 1.4 & 1.7 & -24 & 1.7 & 2.0 & -18 \\
      & $\tau$ & 0.65 & 0.64 & +1.9 & 1.7 & 2.1 & -23 & 2.4 & 2.7 & -16 \\
      $C_{\tau e}^{S,LL}$ & CP  & 7.3 & 4.8 & +34 & 10 & 5.5 & +44 & 12 & 6.8 & +44 \\
      & $\tau$ & 7.8 & 5.7 & +27 & 9.6 & 6.5 & +32 & 14 & 9.2 & +32 \\
      $C_{\tau e}^{S,RR}$ & CP  & 1.3 & 1.3 & +0.99 & 2.4 & 2.4 & +3.0 & 3.0 & 2.9 & +2.9 \\
      & $\tau$ & 1.5 & 1.5 & +0.55 & 2.7 & 2.7 & +1.7 & 3.8 & 3.8 & +1.7 \\
      $C_{\tau e}^{T,LL}$ & CP  & 0.67 & 0.44 & +35 & 0.91 & 0.50 & +45 & 1.1 & 0.62 & +45 \\
      & $\tau$ & 0.72 & 0.53 & +27 & 0.89 & 0.60 & +32 & 1.3 & 0.85 & +32 \\
      $C_{\tau e}^{T,RR}$ & CP  & 0.12 & 0.12 & +1.1 & 0.21 & 0.21 & +2.9 & 0.26 & 0.26 & +2.8 \\
      & $\tau$ & 0.14 & 0.14 & +0.59 & 0.25 & 0.24 & +1.7 & 0.35 & 0.34 & +1.7 \\
      \hline

      $C_{\tau\mu}^{V,LL}$ & CP  & 0.36 & 0.36 & +0.78 & 0.90 & 0.86 & +4.7 & 1.1 & 1.0 & +4.2 \\
      & $\tau$ & 0.33 & 0.33 & +0.12 & 0.74 & 0.76 & -1.8 & 1.0 & 1.0 & -1.6 \\
      $C_{\tau\mu}^{V,LR}$ & CP  & 0.39 & 0.39 & -0.15 & 0.61 & 0.62 & -1.7 & 0.75 & 0.76 & -1.6 \\
      & $\tau$ & 0.26 & 0.26 & -0.15 & 0.32 & 0.32 & -0.31 & 0.45 & 0.45 & -0.31 \\
      $C_{\tau\mu}^{S,LL}$ & CP  & 1.6 & 1.6 & -0.40 & 2.0 & 2.0 & -0.50 & 2.4 & 2.4 & -0.50 \\
      & $\tau$ & 0.94 & 0.94 & -0.17 & 1.1 & 1.1 & -0.17 & 1.5 & 1.5 & -0.18 \\
      $C_{\tau\mu}^{S,RR}$ & CP  & 1.3 & 1.3 & +0.99 & 2.4 & 2.4 & +3.0 & 3.0 & 2.9 & +2.9 \\
      & $\tau$ & 1.5 & 1.5 & +0.55 & 2.7 & 2.7 & +1.7 & 3.8 & 3.8 & +1.7 \\
      $C_{\tau\mu}^{T,LL}$ & CP  & 0.15 & 0.15 & -0.34 & 0.18 & 0.18 & -0.40 & 0.22 & 0.22 & -0.41 \\
      & $\tau$ & 0.087 & 0.087 & -0.11 & 0.10 & 0.10 & -0.20 & 0.14 & 0.14 & -0.14 \\
      $C_{\tau\mu}^{T,RR}$ & CP  & 0.12 & 0.12 & +1.1 & 0.21 & 0.21 & +2.9 & 0.26 & 0.26 & +2.8 \\
      & $\tau$ & 0.14 & 0.14 & +0.59 & 0.25 & 0.24 & +1.7 & 0.35 & 0.34 & +1.7 \\
      \hline

      $C_{\tau\tau}^{V,LL}$ & CP  & 0.34 & 0.33 & +0.30 & 1.3 & 1.4 & -1.6 & 1.3 & 1.4 & -0.83 \\
      & $\tau$ & 0.38 & 0.38 & +0.16 & 1.7 & 1.9 & -6.1 & 1.8 & 1.9 & -1.9 \\
      $C_{\tau\tau}^{V,LR}$ & CP  & 0.31 & 0.31 & -1.3 & 0.40 & 0.41 & -2.6 & 0.49 & 0.51 & -2.6 \\
      & $\tau$ & 0.35 & 0.35 & -0.78 & 0.45 & 0.45 & -1.5 & 0.63 & 0.64 & -1.5 \\
      $C_{\tau\tau}^{S,LL}$ & CP  & 1.2 & 1.2 & -1.5 & 1.4 & 1.4 & -1.9 & 1.7 & 1.7 & -1.9 \\
      & $\tau$ & 1.3 & 1.3 & -0.90 & 1.5 & 1.5 & -1.1 & 2.1 & 2.1 & -1.1 \\
      $C_{\tau\tau}^{S,RR}$ & CP  & 1.3 & 1.3 & +0.99 & 2.4 & 2.4 & +3.0 & 3.0 & 2.9 & +2.9 \\
      & $\tau$ & 1.5 & 1.5 & +0.55 & 2.7 & 2.7 & +1.7 & 3.8 & 3.8 & +1.7 \\
      $C_{\tau\tau}^{T,LL}$ & CP  & 0.11 & 0.11 & -1.5 & 0.12 & 0.12 & -1.9 & 0.15 & 0.15 & -1.9 \\
      & $\tau$ & 0.12 & 0.12 & -0.92 & 0.14 & 0.14 & -1.2 & 0.19 & 0.20 & -1.1 \\
      $C_{\tau\tau}^{T,RR}$ & CP  & 0.12 & 0.12 & +1.1 & 0.21 & 0.21 & +2.9 & 0.26 & 0.26 & +2.8 \\
      & $\tau$ & 0.14 & 0.14 & +0.59 & 0.25 & 0.24 & +1.7 & 0.35 & 0.34 & +1.7 \\
    \end{tabular}
  \end{ruledtabular}
\end{table*}

\begin{table*}[t]

  \renewcommand{\arraystretch}{1.6}
  \caption{Projected 90\% C.L. bounds on the couplings $C_{ab}^X$ [Eq.~\eqref{eq:L6D}] at the ND of a DUNE-like experiment (6.5 years each in $\nu$ and $\bar{\nu}$ modes), assuming only one non-zero coupling at a time. Due to the short baseline, the ND is insensitive to matter effects, rendering the NO and IO limits identical. Dashes (--) denote cases where no contributing processes are generated.}
  \label{tab:Chi2_D_ND}

  \begin{ruledtabular}
    \begin{tabular}{l cc cc cc}
      & \multicolumn{2}{c}{$(\sigma_\epsilon, \sigma_\omega) = (0, 0)$} & \multicolumn{2}{c}{$(\sigma_\epsilon, \sigma_\omega) = (10\%, 0)$} & \multicolumn{2}{c}{$(\sigma_\epsilon, \sigma_\omega) = (10\%, 2\%)$} \\
      \cmidrule(lr){2-3} \cmidrule(lr){4-5} \cmidrule(lr){6-7}
      Coupling & CP & $\tau$ & CP & $\tau$ & CP & $\tau$ \\
      \hline

      $C_{e\mu}^{V,LL}$ & 0.041 & 0.033 & 0.050 & 0.040 & 0.54 & 0.54 \\
      $C_{e\mu}^{V,LR}$ & 0.026 & 0.021 & 0.031 & 0.024 & 0.33 & 0.33 \\
      $C_{e\mu}^{S,LL}$ & 0.093 & 0.073 & 0.11 & 0.085 & 1.2 & 1.2 \\
      $C_{e\mu}^{S,RR}$ & -- & -- & -- & -- & -- & -- \\
      $C_{e\mu}^{T,LL}$ & 0.0084 & 0.0068 & 0.0098 & 0.0078 & 0.11 & 0.11 \\
      $C_{e\mu}^{T,RR}$ & -- & -- & -- & -- & -- & -- \\
      \hline

      $C_{\mu e}^{V,LL}$ & 0.029 & 0.023 & 0.046 & 0.037 & 0.50 & 0.50 \\
      $C_{\mu e}^{V,LR}$ & 0.045 & 0.036 & 0.086 & 0.070 & 0.94 & 0.96 \\
      $C_{\mu e}^{S,LL}$ & -- & -- & -- & -- & -- & -- \\
      $C_{\mu e}^{S,RR}$ & 0.10 & 0.078 & 0.17 & 0.13 & 1.9 & 1.8 \\
      $C_{\mu e}^{T,LL}$ & -- & -- & -- & -- & -- & -- \\
      $C_{\mu e}^{T,RR}$ & 0.0091 & 0.0072 & 0.015 & 0.012 & 0.17 & 0.17 \\
      \hline

      $C_{\mu\mu}^{V,LL}$ & 0.026 & 0.020 & 0.12 & 0.097 & 0.99 & 0.98 \\
      $C_{\mu\mu}^{V,LR}$ & 0.024 & 0.019 & 0.033 & 0.026 & 0.36 & 0.35 \\
      $C_{\mu\mu}^{S,LL}$ & 0.093 & 0.073 & 0.11 & 0.085 & 1.2 & 1.2 \\
      $C_{\mu\mu}^{S,RR}$ & 0.10 & 0.078 & 0.17 & 0.13 & 1.9 & 1.8 \\
      $C_{\mu\mu}^{T,LL}$ & 0.0084 & 0.0068 & 0.0098 & 0.0078 & 0.11 & 0.11 \\
      $C_{\mu\mu}^{T,RR}$ & 0.0091 & 0.0072 & 0.015 & 0.012 & 0.17 & 0.17 \\
      \hline

      $C_{\mu\tau}^{V,LL}$ & 0.029 & 0.023 & 0.046 & 0.037 & 0.50 & 0.50 \\
      $C_{\mu\tau}^{V,LR}$ & 0.045 & 0.036 & 0.086 & 0.070 & 0.94 & 0.96 \\
      $C_{\mu\tau}^{S,LL}$ & -- & -- & -- & -- & -- & -- \\
      $C_{\mu\tau}^{S,RR}$ & 0.10 & 0.078 & 0.17 & 0.13 & 1.9 & 1.8 \\
      $C_{\mu\tau}^{T,LL}$ & -- & -- & -- & -- & -- & -- \\
      $C_{\mu\tau}^{T,RR}$ & 0.0091 & 0.0072 & 0.015 & 0.012 & 0.17 & 0.17 \\
      \hline

      $C_{\tau\mu}^{V,LL}$ & 0.041 & 0.033 & 0.050 & 0.040 & 0.54 & 0.54 \\
      $C_{\tau\mu}^{V,LR}$ & 0.026 & 0.021 & 0.031 & 0.024 & 0.33 & 0.33 \\
      $C_{\tau\mu}^{S,LL}$ & 0.093 & 0.073 & 0.11 & 0.085 & 1.2 & 1.2 \\
      $C_{\tau\mu}^{S,RR}$ & -- & -- & -- & -- & -- & -- \\
      $C_{\tau\mu}^{T,LL}$ & 0.0084 & 0.0068 & 0.0098 & 0.0078 & 0.11 & 0.11 \\
      $C_{\tau\mu}^{T,RR}$ & -- & -- & -- & -- & -- & -- \\
    \end{tabular}
  \end{ruledtabular}
\end{table*}

\begin{table*}[t]
  \renewcommand{\arraystretch}{1.3}

  \caption{Projected 90\% C.L. bounds on the Majorana neutrino couplings $\tilde{C}_{ab}^X$ [Eq.~\eqref{eq:M6D}] at the FD of a DUNE-like experiment (6.5 years each in $\nu$ and $\bar{\nu}$ modes), assuming only one non-zero coupling at a time. For the scalar/pseudoscalar combinations, only $\tilde{C}^{S,L}_{ab}$ is explicitly listed for compactness, as $\tilde{C}^{S,R}_{ab} = \tilde{C}^{P,L}_{ab} = \tilde{C}^{P,R}_{ab} = \tilde{C}^{S,L}_{ab}$. }
  \label{tab:Chi2_M_FD}

  \begin{ruledtabular}
    \begin{tabular}{ll ccc ccc ccc}
      & & \multicolumn{3}{c}{$(\sigma_\epsilon, \sigma_\omega) = (0, 0)$} & \multicolumn{3}{c}{$(\sigma_\epsilon, \sigma_\omega) = (10\%, 0)$} & \multicolumn{3}{c}{$(\sigma_\epsilon, \sigma_\omega) = (10\%, 2\%)$} \\
      \cmidrule(lr){3-5} \cmidrule(lr){6-8} \cmidrule(lr){9-11}
      Coupling  & Mode & NO & IO & $R$ (\%) & NO & IO & $R$ (\%) & NO & IO & $R$ (\%) \\
      \hline

      $\tilde{C}_{ee}^{V,L}$ & CP  & -- & -- & -- & -- & -- & -- & -- & -- & -- \\
      & $\tau$ & -- & -- & -- & -- & -- & -- & -- & -- & -- \\
      $\tilde{C}_{ee}^{V,R}$ & CP  & -- & -- & -- & -- & -- & -- & -- & -- & -- \\
      & $\tau$ & -- & -- & -- & -- & -- & -- & -- & -- & -- \\
      $\tilde{C}_{ee}^{A,L}$ & CP  & 1.4 & 1.5 & -9.2 & 2.3 & 4.1 & -78 & 2.8 & 4.7 & -68 \\
      & $\tau$ & 1.8 & 1.7 & +1.5 & 3.6 & 5.9 & -63 & 5.0 & 7.3 & -45 \\
      $\tilde{C}_{ee}^{A,R}$ & CP  & 1.7 & 1.3 & +24 & 4.6 & 1.6 & +65 & 5.3 & 2.0 & +63 \\
      & $\tau$ & 2.0 & 1.5 & +22 & 3.3 & 1.9 & +42 & 4.6 & 2.7 & +41 \\
      $\tilde{C}_{ee}^{S,L}$ & CP  & 4.4 & 4.1 & +7.9 & 11 & 6.3 & +42 & 13 & 7.7 & +40 \\
      & $\tau$ & 5.4 & 4.8 & +12 & 25 & 7.6 & +69 & 26 & 11 & +60 \\
      $\tilde{C}_{ee}^{T,L}$ & CP  & -- & -- & -- & -- & -- & -- & -- & -- & -- \\
      & $\tau$ & -- & -- & -- & -- & -- & -- & -- & -- & -- \\
      $\tilde{C}_{ee}^{T,R}$ & CP  & -- & -- & -- & -- & -- & -- & -- & -- & -- \\
      & $\tau$ & -- & -- & -- & -- & -- & -- & -- & -- & -- \\
      \hline

      $\tilde{C}_{\mu e}^{V,L}$ & CP  & 0.81 & 0.82 & -0.71 & 2.2 & 2.5 & -14 & 2.6 & 2.8 & -11 \\
      & $\tau$ & 0.50 & 0.50 & -0.10 & 1.8 & 1.8 & -3.7 & 2.2 & 2.3 & -2.0 \\
      $\tilde{C}_{\mu e}^{V,R}$ & CP  & 0.82 & 0.81 & +2.0 & 1.2 & 1.1 & +5.2 & 1.5 & 1.4 & +5.1 \\
      & $\tau$ & 0.49 & 0.49 & +0.36 & 0.67 & 0.67 & +0.82 & 0.95 & 0.94 & +0.82 \\
      $\tilde{C}_{\mu e}^{A,L}$ & CP  & 0.81 & 0.82 & -0.71 & 2.2 & 2.5 & -14 & 2.6 & 2.8 & -11 \\
      & $\tau$ & 0.50 & 0.50 & -0.10 & 1.8 & 1.8 & -3.7 & 2.2 & 2.3 & -2.0 \\
      $\tilde{C}_{\mu e}^{A,R}$ & CP  & 0.82 & 0.81 & +2.0 & 1.2 & 1.1 & +5.2 & 1.5 & 1.4 & +5.1 \\
      & $\tau$ & 0.49 & 0.49 & +0.36 & 0.67 & 0.67 & +0.82 & 0.95 & 0.94 & +0.82 \\
      $\tilde{C}_{\mu e}^{S,L}$ & CP  & 2.4 & 2.4 & +0.62 & 5.9 & 5.3 & +11 & 7.0 & 6.3 & +9.3 \\
      & $\tau$ & 1.5 & 1.5 & +0.14 & 2.9 & 2.9 & +1.5 & 4.1 & 4.0 & +1.4 \\
      $\tilde{C}_{\mu e}^{T,L}$ & CP  & 0.29 & 0.28 & +2.8 & 0.35 & 0.33 & +3.6 & 0.43 & 0.41 & +3.6 \\
      & $\tau$ & 0.17 & 0.17 & +0.52 & 0.20 & 0.20 & +0.60 & 0.28 & 0.28 & +0.60 \\
      $\tilde{C}_{\mu e}^{T,R}$ & CP  & 0.29 & 0.29 & -1.5 & 0.46 & 0.47 & -3.6 & 0.56 & 0.58 & -3.6 \\
      & $\tau$ & 0.18 & 0.18 & -0.22 & 0.30 & 0.30 & -0.66 & 0.42 & 0.43 & -0.64 \\
      \hline

      $\tilde{C}_{\mu\mu}^{V,L}$ & CP  & -- & -- & -- & -- & -- & -- & -- & -- & -- \\
      & $\tau$ & -- & -- & -- & -- & -- & -- & -- & -- & -- \\
      $\tilde{C}_{\mu\mu}^{V,R}$ & CP  & -- & -- & -- & -- & -- & -- & -- & -- & -- \\
      & $\tau$ & -- & -- & -- & -- & -- & -- & -- & -- & -- \\
      $\tilde{C}_{\mu\mu}^{A,L}$ & CP  & 0.43 & 0.43 & +0.00 & 1.2 & 1.2 & +1.1 & 1.4 & 1.4 & +0.85 \\
      & $\tau$ & 0.25 & 0.25 & -0.12 & 0.92 & 0.92 & -0.030 & 1.1 & 1.1 & -0.090 \\
      $\tilde{C}_{\mu\mu}^{A,R}$ & CP  & 0.42 & 0.43 & -0.26 & 0.60 & 0.61 & -0.56 & 0.74 & 0.75 & -0.55 \\
      & $\tau$ & 0.25 & 0.25 & -0.12 & 0.34 & 0.34 & -0.18 & 0.48 & 0.48 & -0.17 \\
      $\tilde{C}_{\mu\mu}^{S,L}$ & CP  & 1.3 & 1.3 & -0.18 & 2.9 & 2.9 & -1.3 & 3.4 & 3.5 & -1.1 \\
      & $\tau$ & 0.74 & 0.74 & -0.15 & 1.5 & 1.5 & -0.22 & 2.0 & 2.1 & -0.21 \\
      $\tilde{C}_{\mu\mu}^{T,L}$ & CP  & -- & -- & -- & -- & -- & -- & -- & -- & -- \\
      & $\tau$ & -- & -- & -- & -- & -- & -- & -- & -- & -- \\
      $\tilde{C}_{\mu\mu}^{T,R}$ & CP  & -- & -- & -- & -- & -- & -- & -- & -- & -- \\
      & $\tau$ & -- & -- & -- & -- & -- & -- & -- & -- & -- \\
    \end{tabular}
  \end{ruledtabular}
\end{table*}

\begin{table*}[t]
  \addtocounter{table}{-1}

  \renewcommand{\arraystretch}{1.3}

  \caption{Sensitivity bounds on Majorana neutrino couplings $\tilde{C}_{\alpha\beta}^{X}$ at the FD (Continued).}

  \begin{ruledtabular}
    \begin{tabular}{ll ccc ccc ccc}
      & & \multicolumn{3}{c}{$(\sigma_\epsilon, \sigma_\omega) = (0, 0)$} & \multicolumn{3}{c}{$(\sigma_\epsilon, \sigma_\omega) = (10\%, 0)$} & \multicolumn{3}{c}{$(\sigma_\epsilon, \sigma_\omega) = (10\%, 2\%)$} \\
      \cmidrule(lr){3-5} \cmidrule(lr){6-8} \cmidrule(lr){9-11}
      Coupling  & Mode & NO & IO & $R$ (\%) & NO & IO & $R$ (\%) & NO & IO & $R$ (\%) \\
      \hline

      $\tilde{C}_{\tau e}^{V,L}$ & CP  & 0.65 & 0.65 & +0.020 & 2.6 & 2.6 & +0.00 & 2.7 & 2.7 & +0.030 \\
      & $\tau$ & 0.74 & 0.74 & +0.28 & 3.6 & 3.6 & +0.26 & 3.7 & 3.7 & +0.28 \\
      $\tilde{C}_{\tau e}^{V,R}$ & CP  & 0.60 & 0.60 & +0.13 & 0.80 & 0.80 & +0.24 & 0.99 & 0.98 & +0.23 \\
      & $\tau$ & 0.68 & 0.68 & +0.29 & 0.89 & 0.88 & +0.29 & 1.2 & 1.2 & +0.29 \\
      $\tilde{C}_{\tau e}^{A,L}$ & CP  & 0.65 & 0.65 & +0.020 & 2.6 & 2.6 & +0.00 & 2.7 & 2.7 & +0.030 \\
      & $\tau$ & 0.74 & 0.74 & +0.28 & 3.6 & 3.6 & +0.26 & 3.7 & 3.7 & +0.28 \\
      $\tilde{C}_{\tau e}^{A,R}$ & CP  & 0.60 & 0.60 & +0.13 & 0.80 & 0.80 & +0.24 & 0.99 & 0.98 & +0.23 \\
      & $\tau$ & 0.68 & 0.68 & +0.29 & 0.89 & 0.88 & +0.29 & 1.2 & 1.2 & +0.29 \\
      $\tilde{C}_{\tau e}^{S,L}$ & CP  & 1.9 & 1.9 & +0.11 & 3.3 & 3.3 & +0.43 & 4.1 & 4.1 & +0.41 \\
      & $\tau$ & 2.1 & 2.1 & +0.31 & 3.6 & 3.6 & +0.32 & 5.0 & 5.0 & +0.32 \\
      $\tilde{C}_{\tau e}^{T,L}$ & CP  & 0.21 & 0.21 & +0.14 & 0.24 & 0.24 & +0.17 & 0.30 & 0.30 & +0.17 \\
      & $\tau$ & 0.24 & 0.23 & +0.25 & 0.27 & 0.27 & +0.26 & 0.38 & 0.38 & +0.29 \\
      $\tilde{C}_{\tau e}^{T,R}$ & CP  & 0.24 & 0.24 & -0.040 & 0.41 & 0.41 & -0.20 & 0.50 & 0.50 & -0.18 \\
      & $\tau$ & 0.27 & 0.27 & +0.30 & 0.48 & 0.48 & +0.31 & 0.67 & 0.67 & +0.30 \\
      \hline

      $\tilde{C}_{\tau\mu}^{V,L}$ & CP  & 0.53 & 0.53 & +0.23 & 2.0 & 1.9 & +5.7 & 2.1 & 2.0 & +2.9 \\
      & $\tau$ & 0.42 & 0.42 & -0.020 & 1.8 & 1.8 & +2.7 & 2.0 & 2.0 & +1.0 \\
      $\tilde{C}_{\tau\mu}^{V,R}$ & CP  & 0.50 & 0.50 & -0.91 & 0.67 & 0.68 & -2.0 & 0.82 & 0.84 & -2.0 \\
      & $\tau$ & 0.40 & 0.41 & -0.35 & 0.54 & 0.54 & -0.65 & 0.76 & 0.76 & -0.65 \\
      $\tilde{C}_{\tau\mu}^{A,L}$ & CP  & 0.53 & 0.53 & +0.23 & 2.0 & 1.9 & +5.7 & 2.1 & 2.0 & +2.9 \\
      & $\tau$ & 0.42 & 0.42 & -0.020 & 1.8 & 1.8 & +2.7 & 2.0 & 2.0 & +1.0 \\
      $\tilde{C}_{\tau\mu}^{A,R}$ & CP  & 0.50 & 0.50 & -0.91 & 0.67 & 0.68 & -2.0 & 0.82 & 0.84 & -2.0 \\
      & $\tau$ & 0.40 & 0.41 & -0.35 & 0.54 & 0.54 & -0.65 & 0.76 & 0.76 & -0.65 \\
      $\tilde{C}_{\tau\mu}^{S,L}$ & CP  & 1.5 & 1.5 & -0.40 & 2.9 & 3.0 & -3.6 & 3.5 & 3.6 & -3.4 \\
      & $\tau$ & 1.2 & 1.2 & -0.21 & 2.3 & 2.3 & -1.1 & 3.2 & 3.2 & -1.0 \\
      $\tilde{C}_{\tau\mu}^{T,L}$ & CP  & 0.17 & 0.17 & -1.1 & 0.20 & 0.20 & -1.4 & 0.25 & 0.25 & -1.4 \\
      & $\tau$ & 0.14 & 0.14 & -0.43 & 0.16 & 0.16 & -0.49 & 0.23 & 0.23 & -0.52 \\
      $\tilde{C}_{\tau\mu}^{T,R}$ & CP  & 0.19 & 0.19 & +0.69 & 0.32 & 0.32 & +1.8 & 0.40 & 0.39 & +1.7 \\
      & $\tau$ & 0.15 & 0.15 & +0.070 & 0.26 & 0.26 & +0.38 & 0.36 & 0.36 & +0.41 \\
      \hline

      $\tilde{C}_{\tau\tau}^{V,L}$ & CP  & -- & -- & -- & -- & -- & -- & -- & -- & -- \\
      & $\tau$ & -- & -- & -- & -- & -- & -- & -- & -- & -- \\
      $\tilde{C}_{\tau\tau}^{V,R}$ & CP  & -- & -- & -- & -- & -- & -- & -- & -- & -- \\
      & $\tau$ & -- & -- & -- & -- & -- & -- & -- & -- & -- \\
      $\tilde{C}_{\tau\tau}^{A,L}$ & CP  & 0.34 & 0.33 & +0.30 & 1.3 & 1.4 & -1.6 & 1.3 & 1.4 & -0.83 \\
      & $\tau$ & 0.38 & 0.38 & +0.16 & 1.7 & 1.9 & -6.1 & 1.8 & 1.9 & -1.9 \\
      $\tilde{C}_{\tau\tau}^{A,R}$ & CP  & 0.31 & 0.31 & -1.3 & 0.40 & 0.41 & -2.6 & 0.49 & 0.51 & -2.6 \\
      & $\tau$ & 0.35 & 0.35 & -0.78 & 0.45 & 0.45 & -1.5 & 0.63 & 0.64 & -1.5 \\
      $\tilde{C}_{\tau\tau}^{S,L}$ & CP  & 0.96 & 0.97 & -0.58 & 1.6 & 1.7 & -4.4 & 2.0 & 2.1 & -4.2 \\
      & $\tau$ & 1.1 & 1.1 & -0.36 & 1.8 & 1.8 & -2.4 & 2.5 & 2.6 & -2.4 \\
      $\tilde{C}_{\tau\tau}^{T,L}$ & CP  & -- & -- & -- & -- & -- & -- & -- & -- & -- \\
      & $\tau$ & -- & -- & -- & -- & -- & -- & -- & -- & -- \\
      $\tilde{C}_{\tau\tau}^{T,R}$ & CP  & -- & -- & -- & -- & -- & -- & -- & -- & -- \\
      & $\tau$ & -- & -- & -- & -- & -- & -- & -- & -- & -- \\
    \end{tabular}
  \end{ruledtabular}
\end{table*}

\begin{table*}[t]
  \renewcommand{\arraystretch}{1.7}
  \caption{Projected 90\% C.L. bounds on the Majorana neutrino couplings $\tilde{C}_{ab}^X$ [Eq.~\eqref{eq:M6D}] at the ND of a DUNE-like experiment (6.5 years each in $\nu$ and $\bar{\nu}$ modes), assuming only one non-zero coupling at a time.}
  \label{tab:Chi2_M_ND}

  \begin{ruledtabular}
    \begin{tabular}{l cc cc cc}
      & \multicolumn{2}{c}{$(\sigma_\epsilon, \sigma_\omega) = (0, 0)$} & \multicolumn{2}{c}{$(\sigma_\epsilon, \sigma_\omega) = (10\%, 0)$} & \multicolumn{2}{c}{$(\sigma_\epsilon, \sigma_\omega) = (10\%, 2\%)$} \\
      \cmidrule(lr){2-3} \cmidrule(lr){4-5} \cmidrule(lr){6-7}
      Coupling  & CP & $\tau$ & CP & $\tau$ & CP & $\tau$ \\
      \hline

      $\tilde{C}_{\mu e}^{V,L}$ & 0.051 & 0.041 & 0.25  & 0.19  & 2.0 & 2.0 \\
      $\tilde{C}_{\mu e}^{V,R}$ & 0.048 & 0.039 & 0.065 & 0.052 & 0.71 & 0.70 \\
      $\tilde{C}_{\mu e}^{A,L}$ & 0.051 & 0.041 & 0.25  & 0.19  & 2.0 & 2.0 \\
      $\tilde{C}_{\mu e}^{A,R}$ & 0.048 & 0.039 & 0.065 & 0.052 & 0.71 & 0.70 \\
      $\tilde{C}_{\mu e}^{S,L}$ & 0.15  & 0.12  & 0.28  & 0.22  & 3.1 & 3.0 \\
      $\tilde{C}_{\mu e}^{T,L}$ & 0.017 & 0.014 & 0.020 & 0.016 & 0.21 & 0.21 \\
      $\tilde{C}_{\mu e}^{T,R}$ & 0.018 & 0.014 & 0.031 & 0.025 & 0.33 & 0.34 \\
      \hline

      $\tilde{C}_{\mu\mu}^{V,L}$ & --    & --    & --    & --    & --   & --   \\
      $\tilde{C}_{\mu\mu}^{V,R}$ & --    & --    & --    & --    & --   & --   \\
      $\tilde{C}_{\mu\mu}^{A,L}$ & 0.026 & 0.020 & 0.12  & 0.097 & 0.99 & 0.98 \\
      $\tilde{C}_{\mu\mu}^{A,R}$ & 0.024 & 0.019 & 0.033 & 0.026 & 0.36 & 0.35 \\
      $\tilde{C}_{\mu\mu}^{S,L}$ & 0.075 & 0.059 & 0.14  & 0.11  & 1.5  & 1.5  \\
      $\tilde{C}_{\mu\mu}^{T,L}$ & --    & --    & --    & --    & --   & --   \\
      $\tilde{C}_{\mu\mu}^{T,R}$ & --    & --    & --    & --    & --   & --   \\
      \hline

      $\tilde{C}_{\tau\mu}^{V,L}$ & 0.051 & 0.041 & 0.25  & 0.19  & 2.0 & 2.0 \\
      $\tilde{C}_{\tau\mu}^{V,R}$ & 0.048 & 0.039 & 0.065 & 0.052 & 0.71 & 0.70 \\
      $\tilde{C}_{\tau\mu}^{A,L}$ & 0.051 & 0.041 & 0.25  & 0.19  & 2.0 & 2.0 \\
      $\tilde{C}_{\tau\mu}^{A,R}$ & 0.048 & 0.039 & 0.065 & 0.052 & 0.71 & 0.70 \\
      $\tilde{C}_{\tau\mu}^{S,L}$ & 0.15  & 0.12  & 0.28  & 0.22  & 3.1 & 3.0 \\
      $\tilde{C}_{\tau\mu}^{T,L}$ & 0.017 & 0.014 & 0.020 & 0.016 & 0.21 & 0.21 \\
      $\tilde{C}_{\tau\mu}^{T,R}$ & 0.018 & 0.014 & 0.031 & 0.025 & 0.33 & 0.34 \\
    \end{tabular}
  \end{ruledtabular}
\end{table*}

\begin{figure*}[phtb]
  \centering
  \includegraphics[width=0.70\textwidth]{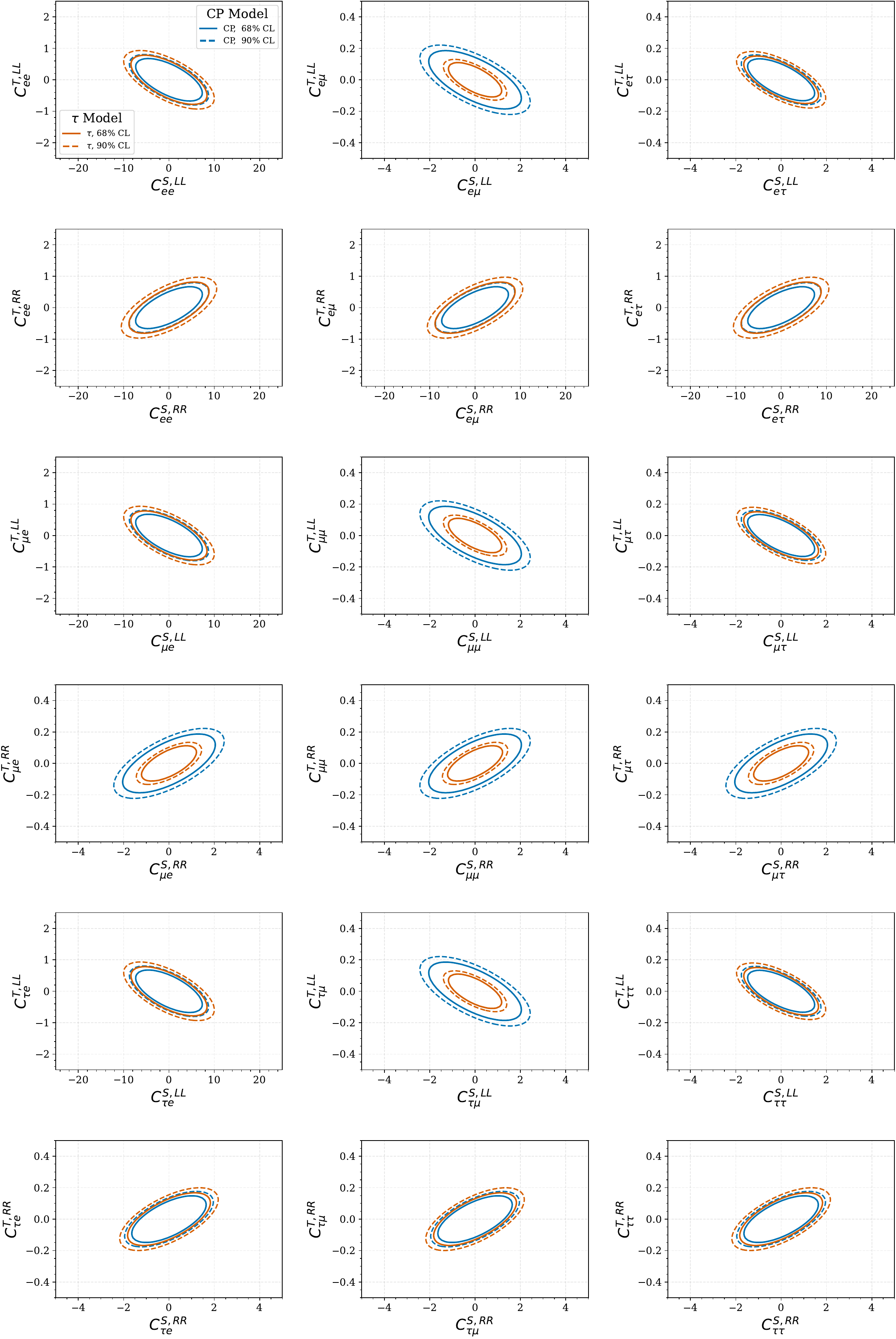}
  \caption{Projected allowed regions illustrating the correlations between Dirac scalar and tensor coefficients at the FD, assuming NO and $(\sigma_\epsilon,\sigma_\omega)=(0,0)$. Blue (orange) contours correspond to the CP-optimized ($\tau$-optimized) beam; solid and dashed curves denote the 68\% and 90\% C.L. regions, respectively. The IO contours are similar and are thus not shown.}
  \label{fig:Chi2_2D_FD_NO}
\end{figure*}

\begin{figure*}[phtb]
  \centering
  \includegraphics[width=0.7\textwidth]{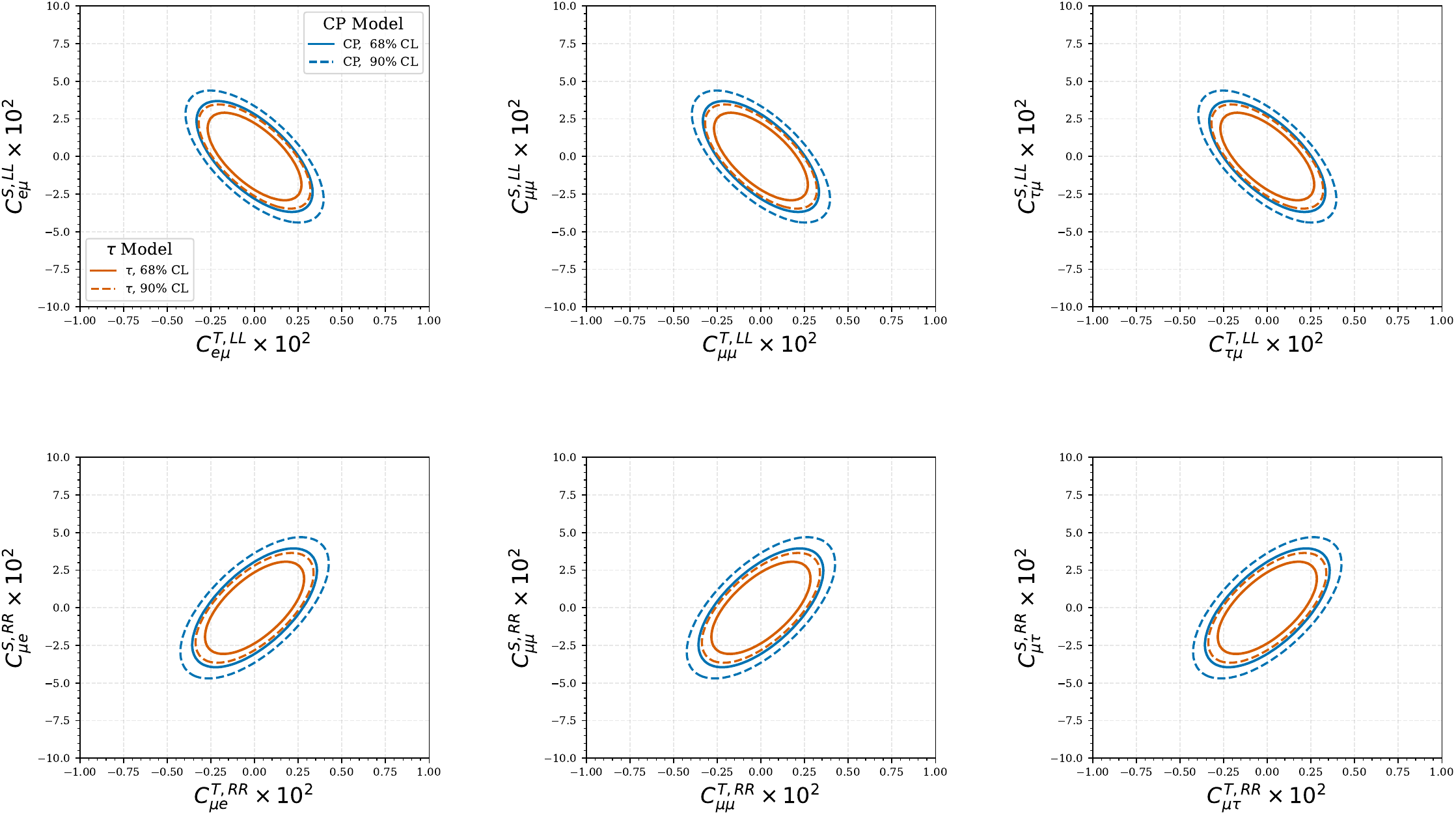}
  \caption{Correlations between selected Dirac scalar and tensor coefficients at the ND. Other conventions follow Fig.~\ref{fig:Chi2_2D_FD_NO}.}
  \label{fig:Chi2_2D_ND_NO}
\end{figure*}

\begin{figure*}[phtb]
  \centering
  \includegraphics[width=0.7\textwidth]{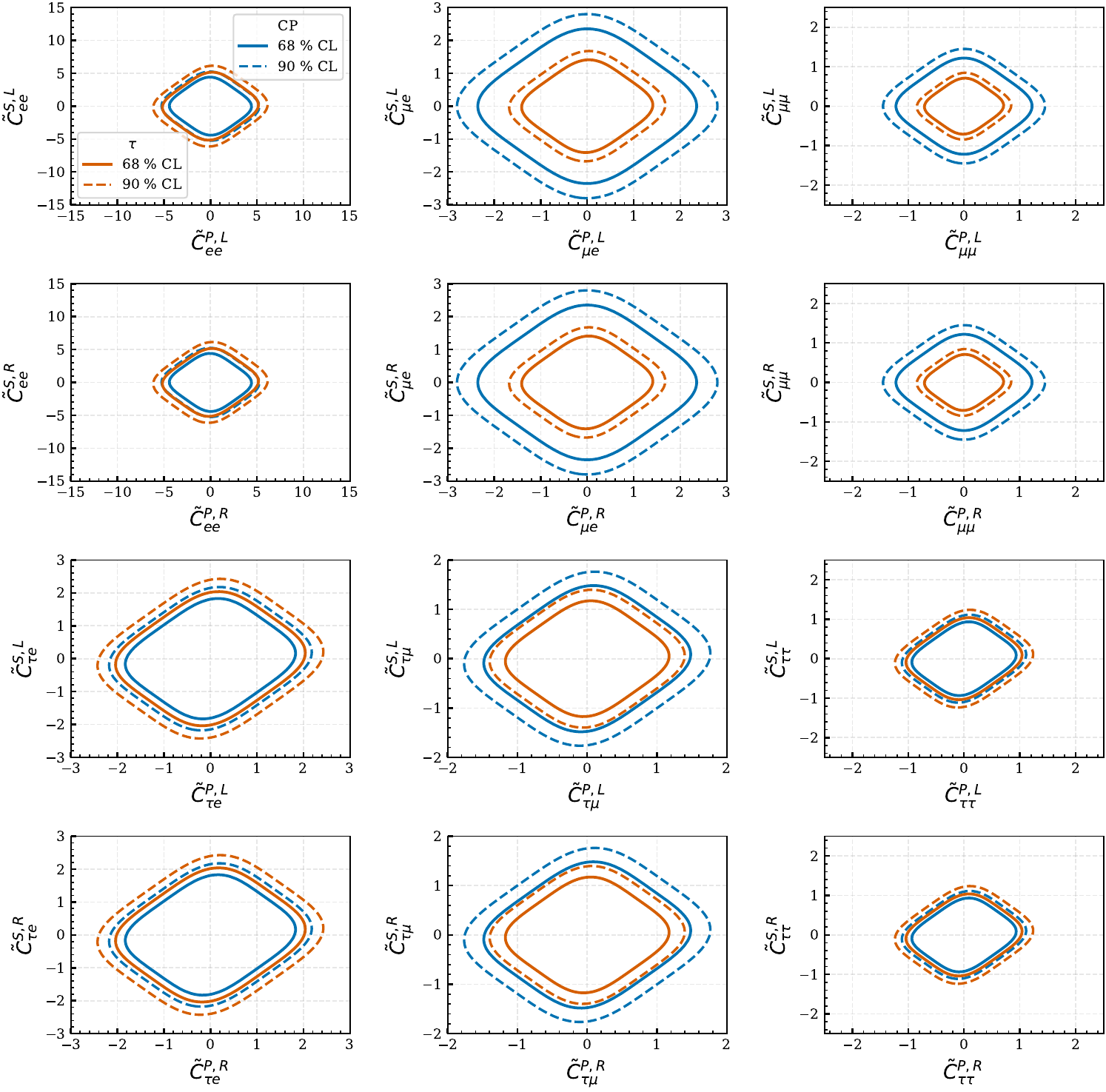}
  \caption{Correlations between selected Majorana scalar and pseudoscalar coefficients at the FD, assuming NO, where $\tilde{C}_{ab}^{P,L/R}$ represents the imaginary part $\mathrm{Im}(\tilde{C}_{ab}^{P,L/R})$.  Other conventions follow Fig.~\ref{fig:Chi2_2D_FD_NO}.}
  \label{fig:Chi2_M_2D_FD_SP}
\end{figure*}

\begin{figure*}[phtb]
  \centering
  \includegraphics[width=0.7\textwidth]{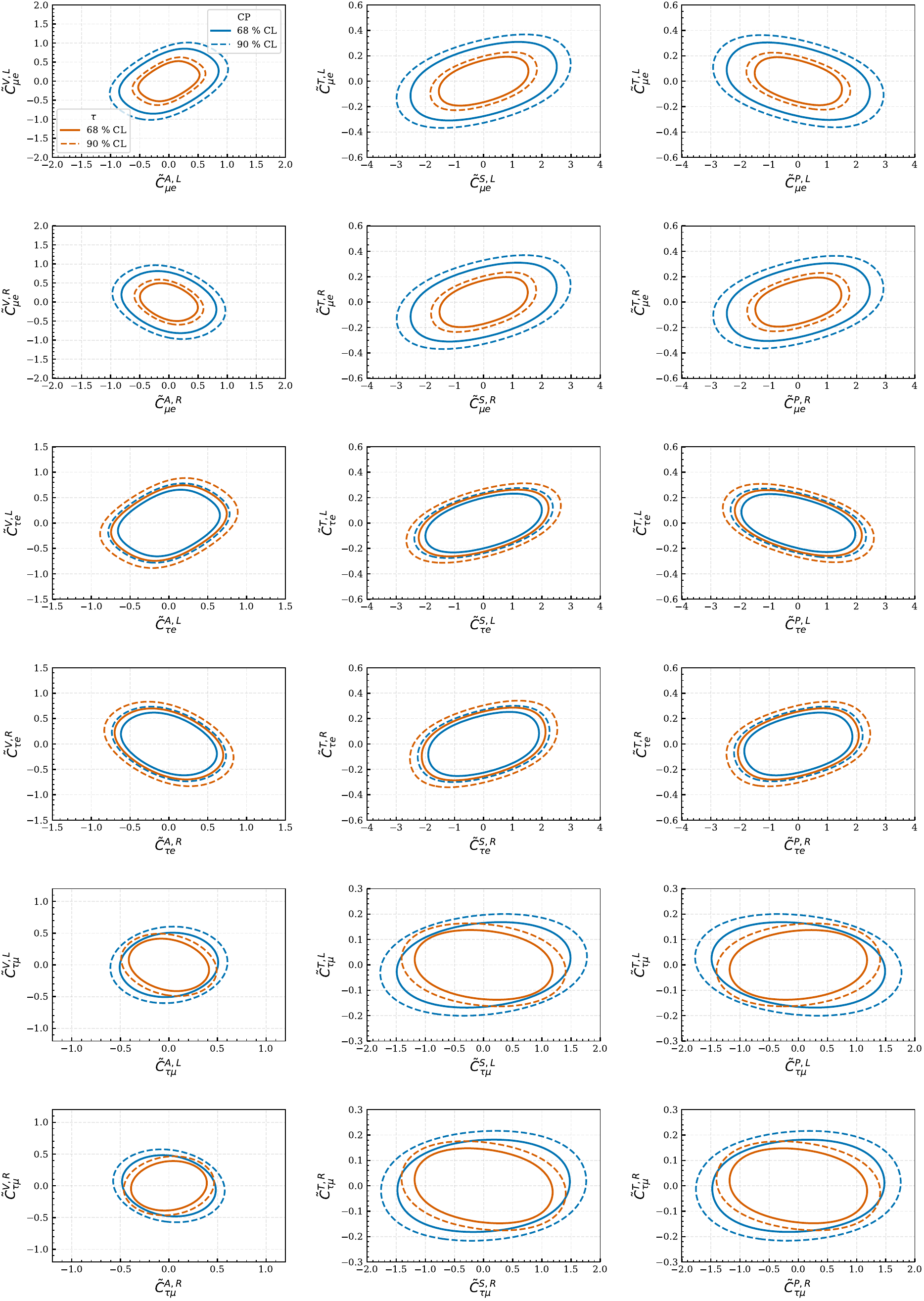}
  \caption{Correlations between selected Majorana vector--axial-vector and scalar/pseudoscalar--tensor coefficient pairs at the FD, assuming NO. Other conventions follow Fig.~\ref{fig:Chi2_2D_FD_NO}.}
  \label{fig:Chi2_M_2D_FD_other}
\end{figure*}

\begin{figure*}[phtb]
  \centering
  \includegraphics[width=0.7\textwidth]{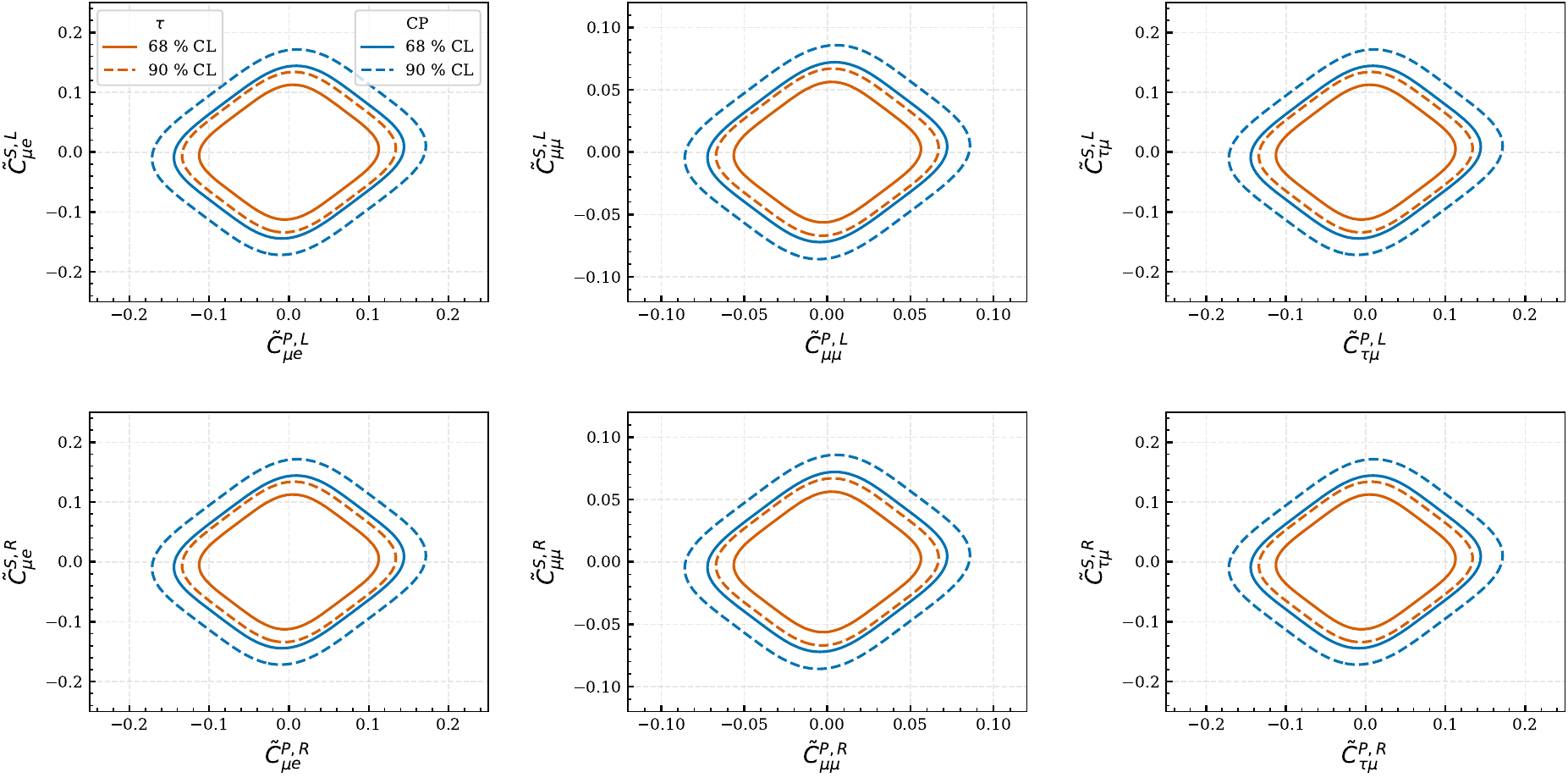}
  \caption{Correlations between selected Majorana scalar and pseudoscalar coefficients at the ND. Conventions follow Fig.~\ref{fig:Chi2_2D_FD_NO} and the neutrino mass ordering is irrelevant at the ND.}
  \label{fig:Chi2_M_2D_ND_SP}
\end{figure*}

\begin{figure*}[phtb]
  \centering
  \includegraphics[width=0.7\textwidth]{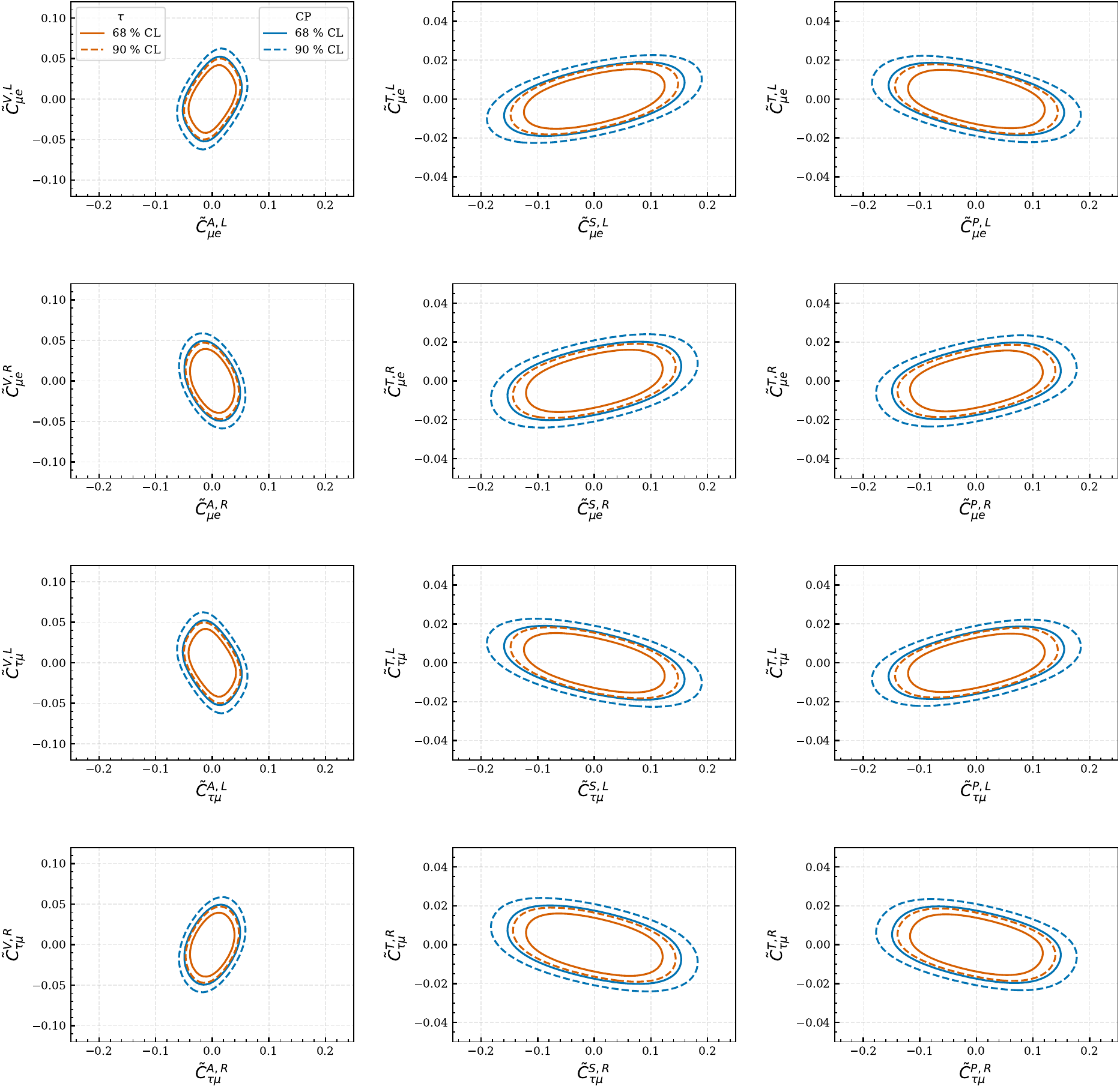}
  \caption{Correlations between selected Majorana vector--axial-vector and scalar/pseudoscalar--tensor coefficient pairs at the ND. Conventions follow Fig.~\ref{fig:Chi2_2D_FD_NO} and the neutrino mass ordering is irrelevant at the ND.}
  \label{fig:Chi2_M_2D_ND_other}
\end{figure*}

\subsection{Operator Correlations}

We now examine correlations between pairs of Wilson coefficients by varying two
coefficients simultaneously while setting all others to zero, focusing on
combinations with nonvanishing interference in the cross sections. For simplicity, all Dirac coefficients are taken to be real. For Majorana neutrinos, all coefficients are likewise assumed real except the pseudoscalar ones, which are taken to be purely imaginary, $\tilde{C}_{ij}^{P,L/R} = i\,\mathrm{Im}\,\tilde{C}_{ij}^{P,L/R}$. The resulting $68\%$ and $90\%$ C.L. contours for representative flavor channels are shown in Figs.~\ref{fig:Chi2_2D_FD_NO}--\ref{fig:Chi2_M_2D_ND_other} for the CP- and $\tau$-optimized fluxes in the idealized $(\sigma_\epsilon,\sigma_\omega)=(0,0)$ scenario. The FD results are shown assuming NO; the corresponding IO contours are similar and thus not displayed, while the neutrino mass ordering is irrelevant at the ND.

For Dirac neutrinos (Figs.~\ref{fig:Chi2_2D_FD_NO} and \ref{fig:Chi2_2D_ND_NO}), the scalar--tensor contours are elliptical, and their major-axis slopes depend on chirality. The coefficient pairs $(C_{ab}^{S,LL},C_{ab}^{T,LL})$ exhibit a negative slope, whereas the $(C_{ab}^{S,RR},C_{ab}^{T,RR})$ pairs exhibit a positive slope. This difference originates from the scalar--tensor interference term in Eq.~\eqref{eq:total_cross_section}. For the $LL$ pair, the PDF difference $f_1^i-f_2^i$ is $f^s-f^{\bar d}$ for $i=\nu_b$ and $f^{\bar s}-f^d$ for $i=\bar\nu_b$. For the $RR$ pair, it is instead $f^d-f^{\bar s}$ for $i=\nu_a$ and $f^{\bar d}-f^s$ for $i=\bar\nu_a$. After the PDF and flux integrations, these two groups yield opposite signs for the scalar--tensor interference term, resulting in the opposite contour slopes.

For Majorana neutrinos, the scalar--pseudoscalar planes in
Figs.~\ref{fig:Chi2_M_2D_FD_SP} and \ref{fig:Chi2_M_2D_ND_SP} exhibit rounded
diamond-shaped contours. With $\tilde{C}_{ab}^{P,\kappa}=i \mathrm{Im}\,\tilde{C}_{ab}^{P,\kappa}$,
the scalar--pseudoscalar factor
in Eq.~\eqref{eq:M_cro_sec2} becomes
$|\tilde{C}_{ab}^{S,\kappa}-s_h\mathrm{Im}\,\tilde{C}_{ab}^{P,\kappa}|^2$.
The opposite signs for $s_h=\pm1$ produce the rounded diamond contours. The
same pattern occurs at both detectors,
with substantially smaller regions at the ND.

The remaining Majorana correlations are shown in
Figs.~\ref{fig:Chi2_M_2D_FD_other} and
\ref{fig:Chi2_M_2D_ND_other}. Their tilt is controlled by both the incoming
chirality, through $s_h$, and whether the incoming flavor is $a$ or $b$,
through $s_\alpha$. The vector--axial-vector amplitudes and the
pseudoscalar--tensor interference contain the product $s_hs_\alpha$;
consequently, their slopes reverse between the left- and right-chiral panels
for a fixed dominant incoming flavor and also reverse when that flavor changes
from $a$ to $b$. The scalar--tensor interference depends on $s_\alpha$ but not
on $s_h$. Its left- and right-chiral contours therefore have the same tilt for
a given flavor composition, while the tilt can reverse between different
flavor channels.

Across most Dirac and Majorana channels, the two beam configurations yield
contours with similar orientations but different allowed areas. The
$\tau$-optimized configuration (orange contours) generally provides tighter
constraints for operator combinations involving the muon flavor. A minor
exception occurs for the $\tau\mu$ channels
at the FD in Fig.~\ref{fig:Chi2_M_2D_FD_other}, where the contours exhibit
slightly different tilt angles.

\section{Summary}
\label{sec:Summary}

In this work, we have investigated the sensitivity of DUNE-like facilities to
$d\leftrightarrow s$ FCNC interactions through inclusive neutral-current DIS.
Within a low-energy effective field theory for both Dirac and Majorana
neutrinos, we have derived the corresponding DIS cross sections and event
yields, incorporated neutrino flavor evolution to the far detector, and
evaluated both the CP- and $\tau$-optimized beam configurations.

Varying one Wilson coefficient at a time in the idealized benchmark scenario
with the $\tau$-optimized flux, the strongest ND bounds on representative
Dirac tensor, vector, and scalar coefficients reach $0.0068$, $0.019$, and
$0.073$, respectively. For Majorana neutrinos, the corresponding tensor,
axial-vector, and scalar/pseudoscalar bounds are $0.014$, $0.019$, and $0.059$.
Due to lower event statistics, the best FD sensitivities are roughly one
order of magnitude weaker, reaching
$\mathcal{O}(10^{-1})\text{--}\mathcal{O}(1)$. Although these bounds are
weaker than indirect rare-kaon limits, which reach
$\mathcal{O}(10^{-6})\text{--}\mathcal{O}(10^{-4})$, the latter apply to
Wilson coefficients in the neutrino mass basis, whereas our scattering results
probe the flavor basis. Across most channels, tensor interactions are the most
tightly constrained and scalar/pseudoscalar interactions the least; this
hierarchy is opposite to that observed in rare kaon decays.

We have also assessed the impact of systematic uncertainties and oscillation
parameters, finding broadly similar patterns for Dirac and Majorana neutrinos.
At the FD, the signal-rate uncertainty ($\sigma_\epsilon=10\%$) is
the dominant systematic effect, substantially weakening vector coefficients,
as well as axial-vector and certain scalar coefficients in the Majorana case,
while affecting tensor bounds more moderately. A background-rate uncertainty of
$\sigma_\omega=2\%$ causes only mild further weakening. At the ND, however,
the bounds are highly sensitive to the background-rate uncertainty, weakening by factors
of roughly $10\text{--}50$ once $\sigma_\omega=2\%$ is included. At
the FD, dependence on the neutrino mass ordering and the CP-violating phase
$\delta_{\mathrm{CP}}$ is largely confined to Wilson coefficients involving
the $e$ flavor. For $(\sigma_\epsilon,\sigma_\omega)=(10\%,0)$,
relative differences between NO and IO can reach approximately $80\%$ with the
CP-optimized flux, while $1\sigma$ variations in $\delta_{\mathrm{CP}}$ can
shift selected electron-flavor limits by about $70\%$ with the CP-optimized
flux and $90\%$ with the $\tau$-optimized flux. By contrast, couplings
involving exclusively $\mu$ and $\tau$ flavors generally exhibit only mild
dependence on the mass ordering and $\delta_{\mathrm{CP}}$.

Finally, we have examined correlations between pairs of Wilson coefficients
with nonvanishing interference. Taking all coefficients to be real except the
Majorana pseudoscalar coefficients, which are purely imaginary, we find
elliptical Dirac scalar--tensor contours, diamond-shaped Majorana
scalar--pseudoscalar regions, and chirality-dependent slope reversals in the
Majorana vector--axial-vector and pseudoscalar--tensor planes.

In conclusion, inclusive neutral-current DIS measurements at future
high-intensity long-baseline neutrino experiments can provide a direct and
complementary probe of effective $d\leftrightarrow s$ FCNC interactions in the
neutrino flavor basis across a broad range of Lorentz and chiral structures.
Beyond long-baseline facilities, the framework developed
here can also be naturally extended to high-energy neutrino scattering at the
forward-physics experiments at the Large Hadron Collider, which we leave for
future investigation.
\vspace{-1.5em}

\section*{Acknowledgments}
We thank Laura Fields for the clarification regarding the neutrino fluxes used in this work.
This work is supported by the National Natural Science Foundation of China under Grants No. 12135006 and No. 12275067,
the Natural Science Foundation of Henan Province under Grant No. 242300421390,
the Science and Technology R\&D Program Joint Fund Project of Henan Province under Grant No. 225200810030,
the Science and Technology Innovation Leading Talent Support Program of Henan Province under Grant No. 254000510039,
as well as the National Key R\&D Program of China under Grant No. 2023YFA1606000.

\clearpage
\bibliographystyle{apsrev4-1}
\bibliography{reference}

\end{document}